\documentclass[letterpaper,twocolumn,prl,aps,superscriptaddress,amsmath,amssymb,10pt]{revtex4-1}
\usepackage{mathptmx}
\usepackage[latin9]{inputenc}
\usepackage{color}
\usepackage{verbatim}
\usepackage{graphicx}
\usepackage{esint}
\usepackage{textcomp}
\usepackage{epstopdf}   
\usepackage{xcolor}
\usepackage{soul}
\usepackage{makecell} 
\usepackage{ulem}
\usepackage[unicode=true,
bookmarks=true,bookmarksnumbered=false,bookmarksopen=false,
breaklinks=false,pdfborder={0 0 1},backref=false,colorlinks=true,linkcolor=magenta,urlcolor=blue,citecolor=blue,pdfstartview={FitH},hyperfootnotes=false]
{hyperref}

\makeatletter

\pdfpageheight\paperheight
\pdfpagewidth\paperwidth

\@ifundefined{textcolor}{}{%
	\definecolor{BLACK}{gray}{0}
	\definecolor{WHITE}{gray}{1}
	\definecolor{RED}{rgb}{1,0,0}
	\definecolor{GREEN}{rgb}{0,1,0}
	\definecolor{BLUE}{rgb}{0,0,1}
	\definecolor{CYAN}{cmyk}{1,0,0,0}
	\definecolor{MAGENTA}{cmyk}{0,1,0,0}
	\definecolor{YELLOW}{cmyk}{0,0,1,0}
}

\newcommand{\bra}[1]{\ensuremath{\left\langle#1\right|}}
\newcommand{\ket}[1]{\ensuremath{\left|#1\right\rangle}}

\definecolor{blue}{rgb}{0,0,1}
\definecolor{red}{rgb}{1,0,0}
\definecolor{green}{rgb}{0,1,0}

\makeatother

\begin{document}
	\title{Experimental Demonstration of Entanglement Pumping with Bosonic Logical Qubits}	
	\author{Jie Zhou}
	\thanks{These two authors contributed equally to this work.}
	\author{Chuanlong Ma}
	\thanks{These two authors contributed equally to this work.}
	\affiliation{Center for Quantum Information, Institute for Interdisciplinary Information Sciences, Tsinghua University, Beijing 100084, China}
	
	\author{Yifang Xu}
	\affiliation{Center for Quantum Information, Institute for Interdisciplinary Information Sciences, Tsinghua University, Beijing 100084, China}
	
	\author{Weizhou Cai}
	\affiliation{CAS Key Laboratory of Quantum Information, University of Science and Technology of China, Hefei 230026, China}
	\author{Hongwei Huang}
	\affiliation{Center for Quantum Information, Institute for Interdisciplinary Information Sciences, Tsinghua University, Beijing 100084, China}	
	\author{Lida Sun}
	\affiliation{Center for Quantum Information, Institute for Interdisciplinary Information Sciences, Tsinghua University, Beijing 100084, China}	
	\author{Guangming Xue}
	\affiliation{Beijing Academy of Quantum Information Sciences, Beijing, China}
	\affiliation{Hefei National Laboratory, Hefei 230088, China}
	\author{Ziyue Hua}
	\affiliation{Center for Quantum Information, Institute for Interdisciplinary Information Sciences, Tsinghua University, Beijing 100084, China}
	\author{Haifeng Yu}
	\affiliation{Beijing Academy of Quantum Information Sciences, Beijing, China}
	\affiliation{Hefei National Laboratory, Hefei 230088, China}
	
	\author{Weiting Wang}
	\email{wangwt2020@tsinghua.edu.cn}
	\affiliation{Center for Quantum Information, Institute for Interdisciplinary Information Sciences, Tsinghua University, Beijing 100084, China}
	
	\author{Chang-Ling Zou}
	\email{clzou321@ustc.edu.cn}
	\affiliation{CAS Key Laboratory of Quantum Information, University of Science and Technology of China, Hefei 230026, China}
	\affiliation{Hefei National Laboratory, Hefei 230088, China}

	\author{Luyan Sun}
	\email{luyansun@tsinghua.edu.cn}
	\affiliation{Center for Quantum Information, Institute for Interdisciplinary Information Sciences, Tsinghua University, Beijing 100084, China}
	\affiliation{Hefei National Laboratory, Hefei 230088, China}
	
	\begin{abstract}
		Entanglement is crucial for quantum networks and computation, yet maintaining high-fidelity entangled quantum states is hindered by decoherence and resource-intensive purification methods. Here, we experimentally demonstrate entanglement pumping, utilizing bosonic quantum error correction (QEC) codes as long-coherence-time storage qubits. By repetitively generating entanglement with short-coherence-time qubits and injecting it into {error-detectable logical qubits}, our approach effectively preserves entanglement. Through error-detection to discard error states and entanglement pumping to mitigate errors within the code space, we extend  {the existence time of entanglement} by nearly 50\% compared to the case without entanglement pumping. {This entanglement pumping scheme can additionally serve as an erasure detection protocol for the dual-rail code.} This work highlights the potential of bosonic logical qubits for scalable quantum networks and introduces a novel paradigm for efficient entanglement management.
	\end{abstract}
	
	\maketitle
	
	\textit{Introduction--} Entanglement plays a pivotal role in quantum physics, not only enhancing our understanding of quantum mechanics~\cite{Einstein1935,Clauser1969,Hill2022} but also serving as a foundational resource for crucial applications such as quantum teleportation~\cite{PhysRevLett.70.1895,Bouwmeester1997,Hu2023}, quantum sensing \cite{Degen2017,Pirandola2018}, and quantum networks \cite{PhysRevLett.78.3221,Kimble2008}. The creation and maintenance of high-fidelity entangled states between nonlocal nodes are of substantial significance for the advancement of these technologies. However, the preparation and lifespan of entangled states are often compromised by factors such as decoherence of qubits, noisy transmission channels, and imperfect measurements and gate operations~\cite{RevModPhys.93.025005,cai2023protecting}. Traditional important methods to enhance entanglement fidelity, such as entanglement purification protocols~\cite{Dur2007,Zhou2016,Yan2023,Hu2021,HUANG2022593,PhysRevA.105.062418}, are highly resource intensive and require the consumption of large numbers of entangled pairs, posing significant experimental challenges. Moreover, interfaces between imperfect entangled pairs and communication channels typically suffer from reduced lifetimes due to the need for rapid interactions and higher entanglement generation rates~\cite{Dur1999,Childress2006}, further limiting the scalability and efficiency of quantum networks.
	
	To address the challenges of resource consumption and entanglement lifetime, innovative strategies are required. One promising approach involves the use of quantum error correction (QEC) codes within local nodes to create long-coherence-time storage qubits~\cite{cai2023protecting,Wang2016,Rosenblum2018,Chou2018,Erhard2021,RyanAnderson2022}. By encoding physical qubits into logical qubits, specific types of errors can be detected and corrected, thereby extending the coherence time and enhancing entanglement fidelity. Additionally, employing entanglement pumping (EP)~\cite{Dur1999,PhysRevLett.128.080504,JIANG2010}, a more efficient form of entanglement purification, allows for the continuous generation of partially entangled pairs, which are then used to reinforce the entanglement between long-lived storage qubits. This approach reduces resource requirements by utilizing a minimal number of qubits to maintain entangled pairs, as opposed to traditional entanglement purification which consumes exponentially many pairs. The combination of QEC-enhanced storage and EP offers a robust solution for sustaining entanglement in quantum networks, facilitating more efficient and scalable quantum communication and computation by continuously injecting entanglement resources and mitigating errors within the code space.
	
	In this Letter, we experimentally demonstrate the first implementation of EP applied to error-detectable logical qubits. By integrating QEC codes as long-coherence-time storage qubits with a repetitive entanglement generation scheme, our EP protocol effectively suppresses decoherence through quantum error detection (ED). Our experimental results show that the application of both EP and ED strategies extends  {the existence time of entanglement} by nearly 50\% compared to using only ED.  These findings underscore the potential of bosonic logical qubits~\cite{cai2020bosonic,Joshi_2021} in quantum networks and introduce EP as a novel and efficient paradigm for entanglement management. Furthermore, the EP protocol can constitute an erasure detection for a dual-rail code~\cite{Koottandavida2024,Chou2024}, and our EP scheme for binomial qubits can be interpreted as a concatenation of binomial codes with dual-rail encoding. This advancement offers significant implications for the development of reliable and scalable quantum communication and distributed computation systems~\cite{Cirac1999,Cacciapuoti2020,PhysRevA.76.062323,PhysRevA.89.022317,PRXQuantum.2.017002}, providing new insights into the optimization of entanglement preservation techniques.
	
	\begin{figure}[ht]
		\centering
		\includegraphics{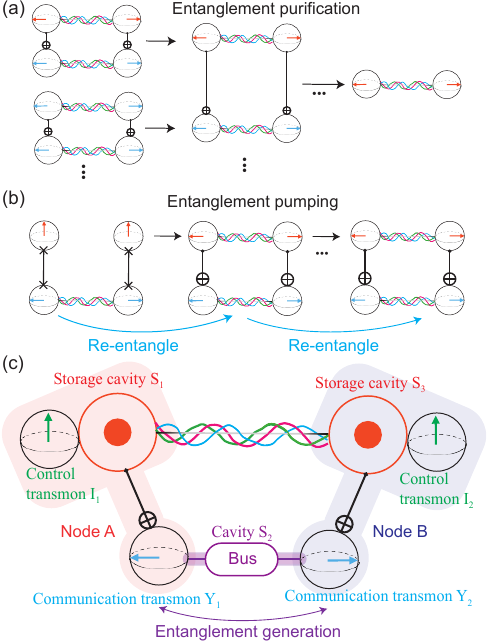}
		\caption{Experimental device and principles for EP. (a) Schematic for entanglement purification. States from previous successful purification serve as inputs for the next round of purification. The total number of pairs increases exponentially with the number of rounds. (b) Schematic for EP. Entangled states are repeatedly generated on one pair of qubits and used to purify another pair. (c) Schematic of the experimental system. It contains three high-quality superconducting cavities and four transmon qubits.}
		\label{fig:fig1}
	\end{figure}

	\textit{Principle and experimental setup--} Figure~\ref{fig:fig1}(a) illustrates the process of entanglement purification. When many copies of entangled states are prepared, we can perform gates between two pairs of qubits, measure one pair, and postselect the remaining pair based on the measurement outcomes. For $2^N$ pairs of qubits with long coherence times, we can produce $2^{N-1}$ pairs by consuming half of the resource of low-quality entangled states. By repeating this process for all remaining pairs of qubits, we eventually obtain one pair of qubits with high-fidelity entanglement, at the cost of sacrificing the quantum resources of many low-fidelity entanglement pairs stored in the coupled qubits. However, this is not easily realized with current quantum systems because the generation of low-quality entanglement pairs is relatively slow, and the available qubits are limited. Therefore, the relevant experimental demonstration remains a challenge.
	
	Figure~\ref{fig:fig1}(b) shows a resource-efficient EP scheme for preparing quantum entangled states with only two pairs of qubits. One pair serves as communication qubits, which have relatively short coherence times but can couple to an intermediate bus waveguide or carriers to generate entanglement. In contrast, the other pair consists of qubits with long coherence times but lacks efficient channels for directly establishing entanglement. The high-quality qubits can serve as memory units, with their entanglement generated and sustained by repetitively pumping entanglement from the communication qubits. Since the communication qubits are reused during the process, there is no need for many pairs of communication qubits or long-coherence-time qubits.
	
	We experimentally demonstrate EP within a superconducting quantum system. {The device used in this experiment is the same as in our previous study~\cite{cai2023protecting}.} The schematic of our experimental device is shown in Fig.~\ref{fig:fig1}(c). Two 3D coaxial cavities ($S_1$ and $S_3$) serve as indirectly connected nodes ($A$ and $B$) in the quantum network, aiming to store entangled quantum states. Each cavity is coupled to a control transmon qubit ($I_1$ and $I_2$, respectively), enabling arbitrary local operations on the cavities through numerically optimized pulses applied to both the cavities and the transmons~\cite{Heeres2017,KHANEJA2005296,Krastanov2015, Reinier2015,Eickbusch2022}. The two cavities are interconnected by two communication transmon qubits ($Y_1$ and $Y_2$) and a bus cavity ($S_2$). With the assistance of bus resonator $S_2$, entangled states can be repeatedly generated on the communication qubits $Y_1$ and $Y_2$. Thus, we can utilize the regenerated entangled pairs to initialize and sustain the entanglement in the two storage cavities ($S_1$ and $S_3$). 
	
	\begin{figure}[ht]
		\includegraphics{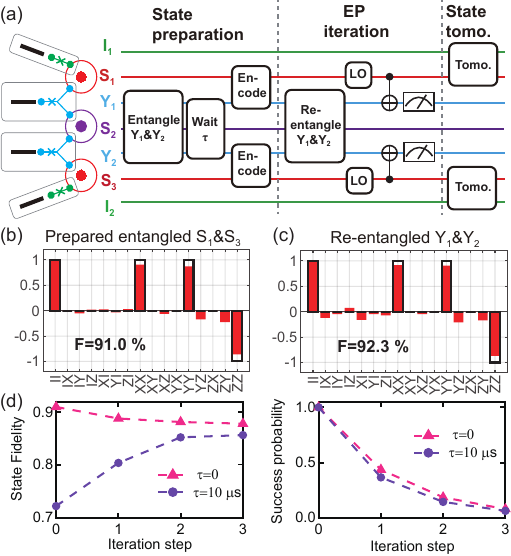}
		\caption{Repetitive EP of physical qubits. (a)  Experimental sequence for repetitive entanglement purification; local operations (LO). (b) Prepared entangled state $\left|\psi\right\rangle=\frac{\left|01\right\rangle+\left|10\right\rangle}{\sqrt{2}}$ in cavities $S_1$ and $S_3$ for $\tau=0$. (c) Reentangled state in communication transmon qubits $Y_1$ and $Y_2$ with an entangled photonic qubit state in cavities $S_1$ and $S_3$. (d) The experimental results of continuous entanglement purification. Three rounds of non-local purification processes are performed.}
		\label{fig:fig2}
	\end{figure}
	
	\textit{EP for the physical qubits--}  We begin by encoding quantum information into the two lowest photon number states, $\ket{0}$ and $\ket{1}$, of the storage cavities. Figure~\ref{fig:fig2}(a) shows the detailed experimental sequence. An initial entangled state $\ket{\psi}_+=\frac{1}{\sqrt2}(\ket{01}+\ket{10})$ is prepared between $S_1$ and $S_3$ by first entangling the communication qubits $Y_1$ and $Y_2$ through $S_2$ to form $\frac{1}{\sqrt2}(\ket{ge}+\ket{eg})$, which is then encoded into the storage cavities via local transmon-cavity swap operations at each node. Here, $\ket{g}$ and $\ket{e}$ represent the ground and excited states of the communication qubit, respectively. The entangled photonic qubits in $S_1$ and $S_3$ are characterized through state tomography using $I_1$ and $I_2$. As shown in Fig.~\ref{fig:fig2}(b), the initial entangled photonic qubits exhibit a state fidelity of $91.0\%$. This entangled state is continuously pumped by repetitively regenerating entangled states in $Y_1$ and $Y_2$ and performing the entanglement purification between the communication and photonic qubits. It is important to consider the dispersive coupling between the storage cavities and communication qubits when optimizing the control pulses for the reentangling gate. The regenerated entanglement between the communication qubits achieves a fidelity of $92.3\%$, as shown in Fig.~\ref{fig:fig2}(c). 
	
	According to the entanglement purification schemes~\cite{bennett1996purification,bennett1996mixed}, bilateral CNOT gates and postselection of the communication qubit measurements in the $\ket{gg}$ state can effectively suppress unwanted components of $\ket{\phi}_\pm=\frac{1}{\sqrt2}(\ket{00}\pm\ket{11})$ in the photonic qubit state. However, errors accumulate in the $\ket{\psi}_-=\frac{1}{\sqrt2}(\ket{01}-\ket{10})$ state over multiple EP rounds, limiting the fidelity of the stored entangled state. To address this issue, we apply local single-qubit operations~\cite{deutsch1996quantum} to exchange the proportion of $\ket{\psi}_-$ with $\ket{\phi}_\pm$ before the CNOT gates. Both communication qubits are measured after the CNOT gates for postselection. Ideally, we should keep the outputs when both communication qubits are in their ground states $\ket{gg}$ or excited states $\ket{ee}$. However, measuring the $\ket{ee}$ state causes an additional phase to the states in the storage cavities. Moreover, we have to reset the communication qubits to $\ket{gg}$ before the next pumping cycle, thus introducing additional errors. For simplicity, we discard $\ket{ee}$ events at the cost of reducing the success rate of purification in each round of EP by approximately half.
	
	Figure~\ref{fig:fig2}(d) presents the evolution of entanglement state fidelity in the storage cavities during the repetitive EP process. When EP is performed directly without any extra wait time ($\tau=0$), we find that the fidelity decreases with successive EP iterations and eventually saturates at a value of 87.8\%. This behavior can be attributed to the accumulation of operation errors during the entanglement purification process and the use of imperfectly regenerated entanglement. Consequently, the steady-state achievable fidelity is limited by a balance of these imperfections, although the fidelity of the regenerated entanglement between the communication qubits is higher. 
	
	In contrast, by introducing a wait time of $\tau=10~\mathrm{\mu s}$ {before} EP iterations, which introduces errors in the storage cavities, we observe that EP can significantly improve the entanglement fidelity. Starting from an initial fidelity of $72.2\%$, the fidelity progressively increases with each EP round and ultimately reaches a saturation value of approximately $85.7\%$. This result confirms that the EP can effectively enhance and sustain improved entanglement fidelity when local memory decoherence is pronounced. Notably, the experiments demonstrate up to four rounds of EP, with the success probability of each round diminishing considerably as the number of iterations increases, revealing the trade-off between fidelity and resources.{When the qubit and cavity lifetimes reach the state-of-the-art, the resulting saturated entanglement fidelity could exceed $99\%$.} Numerical simulations and error budget analyses of the EP process can be found in Supplemental Material~\cite{supplement}.~\nocite{RN9,PhysRevB.79.184301,PhysRevA.40.4277,deFouquieres2011,RN112,RN5,PRXQuantum.4.030336,Dong2015,PhysRevLett.117.250502,PhysRevLett.116.180501}

	\begin{figure}[t]
		\includegraphics{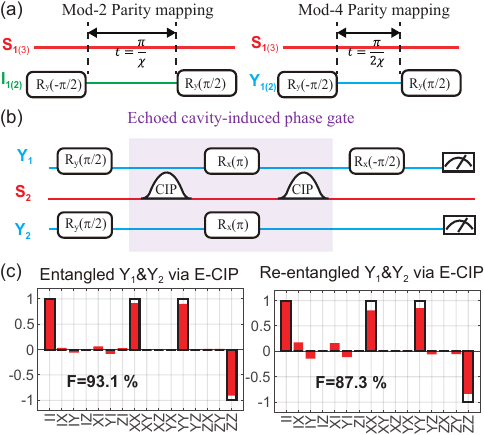}
		\caption{Gate for purification of logical qubits. (a) Diagram for {the regular (mod-2) parity mapping operation and mod-4 parity mapping operation}. (b) Experimental sequence for re-entangling communication qubits $Y_1$ and $Y_2$ using an echoed CIP gate. (c) Generated entangled state without (left) and with (right) a binomially entangled state in cavities $S_1$ and $S_3$.}
		\label{fig:fig3}
	\end{figure}
	
	{\textit{Connections between EP and dual-rail code--} It is noted that the two-cavity entangled state $\ket{\psi}_+$ corresponds to a superposition state of a dual-rail qubit~\cite{Koottandavida2024,Chou2024}.  	In fact, if the local operations are removed, the EP scheme functions as an erasure detection protocol for dual-rail qubits. Applying bilateral CNOT gates to an arbitrary dual-rail qubit state with entangled communication qubits $ (\alpha\ket{01}+\beta\ket{10}) \otimes \frac{1}{\sqrt2}(\ket{ge}+\ket{eg})$, we obtain a final state $ (\alpha\ket{01}+\beta\ket{10}) \otimes \frac{1}{\sqrt2}(\ket{gg}+\ket{ee})$, where $\alpha$ and $\beta$ are arbitrary complex numbers satisfying $|\alpha|^2+|\beta|^2=1$. However, {a single-photon-loss error in either storage cavity collapses the initial state to $\ket{00} \otimes \frac{1}{\sqrt2}(\ket{ge}+\ket{eg})$, marking it as an erasure error}. Applying bilateral CNOT gates to this error state yields the state $\ket{00} \otimes \frac{1}{\sqrt2}(\ket{ge}+\ket{eg})$. Thus, by measuring whether the two communication qubits are in different states, we can determine whether an erasure error occurs. Moreover, replacing the CNOT gates with parity mapping operations shown in the left panel of Fig.~\ref{fig:fig3}(a), we can determine whether the states in the two cavities have identical parity (See supplemental Material~\cite{supplement}).} 
	
	\textit{{EP} for binomial logical qubits--} To achieve longer coherence for stored entanglement, we can encode the states in the cavities using QEC codes. Here, we utilize the lowest-order binomial code to encode the logical qubits~\cite{PhysRevX.6.031006, Hu2019,Ma2020,Ni2023}, where the code space is spanned by $\ket{0}_L=({\ket{0}+\ket{4}})/{\sqrt 2}$ and $\ket{1}_L=\ket{2}$. With this encoding, we can perform error detection for cavities $S_1$ and $S_3$ to mitigate the excitation loss errors in the memory. If a single-photon-loss error is indicated by the output of ED, the entangled logical qubits will be discarded. {Furthermore, by replacing the CNOT gates in the EP scheme with the mod-4 parity mapping shown in the right panel of Fig.~\ref{fig:fig3}(a),  we can detect whether {a logical bit-flip error} or a two-photon-loss error occurs in the cavities.} By combining ED with EP, high-quality entangled-logical-qubit states can be prepared and sustained probabilistically, requiring only local operations on the cavities and control qubits in each node, along with classical communication between nodes.  {Note that this combination can be interpreted as a concatenation of binomial codes with dual-rail encoding~\cite{supplement}.}

	\begin{figure}[t]
		\includegraphics{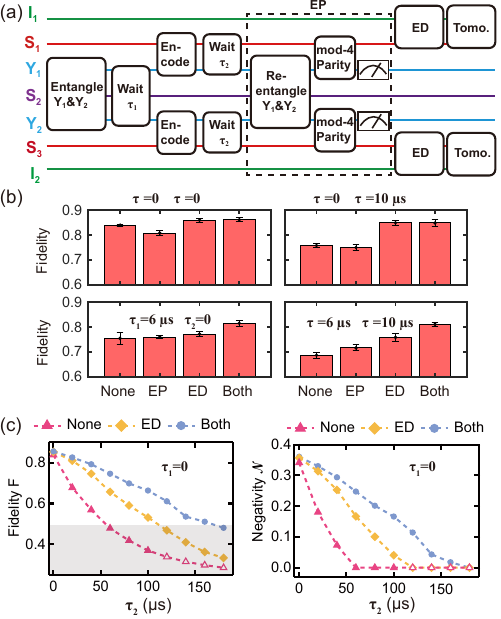}
		\caption{The existence time of entanglement in logical qubits through various purification schemes. (a) Experimental sequence for implementing both EP and ED. Wait times $\tau_1$ and $\tau_2$ are introduced to add logical errors and single-photon-loss errors, respectively. (b) State fidelity for scenarios with no purification, only EP, only ED, and the combination of both ED and EP under different wait times. (c) The influence of different purification schemes on the existence time of entanglement in logical qubits. {The fidelity $F$ and entanglement negativity $\mathcal{N}$ vary with $\tau_2$ for $\tau_1=0$. Values $F>0.5$ (left panel, above shaded region) and $\mathcal{N}>0$ (right panel) confirm the presence of entanglement. The last four points of the ``none" selective curves are from simulations}.}
		\label{fig:fig4}
	\end{figure}    
	
	{For EP of logical qubits, one key challenge arises: the high-fidelity re-entanglement between communication qubits in the presence of multiple photons in the storage cavities.} Considering the dispersive coupling between the communication and logical qubits, we implement an echoed cavity-induced phase (CIP) gate to reentangle the communication qubits. The mechanism of the CIP relies on the cavity-induced geometric phase to the qubits under external driving when they are dispersively coupled to a cavity~\cite{Dong2015,PhysRevLett.116.180501,PhysRevLett.117.250502}.  Ideally, the CIP gate can convert the initial state of the communication qubits $\left|\psi_i\right\rangle=(\left|gg\right\rangle+\left|ge\right\rangle+\left|eg\right\rangle+\left|ee\right\rangle)/2$ to a final state $\left|\psi_f\right\rangle=(\left|ee\right\rangle+e^{i\Phi_1}\left|eg\right\rangle+e^{i\Phi_2}\left|ge\right\rangle+e^{i\Phi_3}\left|ee\right\rangle)/2$, with $\Phi_{1,2,3}$ being the corresponding phases.

	As shown in Fig.~\ref{fig:fig3}(b), by inserting two flip operations $R_x(\pi)$ on the communication qubits between two identical CIP gates, the phases induced by the dispersive couplings with $S_1$ and $S_3$ are canceled out, and the converted final state becomes $\left|\psi_{f2}\right\rangle=(\left|gg\right\rangle+e^{i(\Phi_1+\Phi_2-\Phi_3)}\left|eg\right\rangle+e^{i(\Phi_1+\Phi_2-\Phi_3)}\left|ge\right\rangle+\left|ee\right\rangle)/2$. By carefully adjusting the duration of the CIP gates (see Supplemental Material~\cite{supplement}) , we can satisfy the condition $\Phi_1+\Phi_2-\Phi_3=\pi/2$. Finally, by applying a $R_x(-\frac{\pi}{2})$ rotation to $Y_1$, we obtain the target entangled state $\left|\psi_{f3}\right\rangle=({\left|eg\right\rangle+\left|ge\right\rangle})/{\sqrt{2}}$ on the communication qubits. {We characterize this entangled state generated via echoed CIP gates in experiment, with the results shown in Fig.~\ref{fig:fig3}(c).}
	When both cavities $S_1$ and $S_3$ are in the vacuum state, the fidelity of the entangled communication qubits is 93.1\%. When the cavities are prepared in the entangled binomial logical states, the fidelity decreases to 87.3\%.

	Equipped with the gates designed for logical qubits, we experimentally combine EP and ED methods to protect entanglement between two indirectly connected nodes. The experimental procedure, depicted in Fig.~\ref{fig:fig4}(a), incorporates two relaxation periods: $\tau_1$ before encoding, which allows for the introduction of {errors within the code space}, and $\tau_2$ after encoding, which enables the occurrence of single-photon-loss errors {and two-photon-loss errors.} Figure~\ref{fig:fig4}(b) presents a comprehensive analysis of the fidelity improvements achieved through various EP and ED strategies under different error scenarios ($\tau_{1,2}$). In the absence of intentionally introduced errors $(\tau_1=0, \tau_2=0)$, ED alone improves fidelity, primarily by mitigating single-photon-loss errors from the state preparation process. Conversely, the implementation of EP reduces fidelity. Interestingly, the simultaneous application of ED and EP yields a fidelity comparable to that obtained with ED alone. This can be attributed to the longer gate times associated with EP, which introduce additional single-photon-loss errors that can subsequently be compensated by ED.
	
	When relaxation is introduced only after encoding $(\tau_1=0, \tau_2=\mathrm{10~\mu s})$, single-photon loss becomes the dominant source of error. In this case, the changes in fidelity are similar to those observed when no errors are intentionally introduced. {In contrast, when relaxation is introduced only before encoding $(\tau_1=6~\mathrm{\mu s}, \tau_2=0)$, the decay of the communication qubits leads to $\ket{0_L}\ket{0_L}$ in the cavities, and this logical error can be detected by EP. However, due to single-photon-loss errors and measurement errors introduced during the $5.9~\mu$s EP process, the improvement achieved by EP in this case is not significant in the experiment. We simulate the effect of introducing this relaxation process (characterized by $\tau_1$), as shown in Supplemental Material~\cite{supplement}, and find that the benefit of EP becomes more evident as $\tau_1$ increases. When both EP and ED are applied, EP mitigates logical-space errors, while ED compensates for single-photon-loss errors arising from state preparation and the EP process. Thus, both ED and EP can independently enhance the fidelity by mitigating different types of errors, and their combination leads to an even more pronounced improvement.}

	To evaluate the potential of the concatenated encoding in practical applications, we perform experiments without intentionally introducing relaxation before the encoding stage ($\tau_1=0$). Instead, we focus on extending the relaxation time $\tau_2$ and characterize the preservation of coherence by monitoring {the fidelity $F$ and entanglement negativity $\mathcal{N}$ of the entangled state~\cite{vidal2002computable}. The experimental results are depicted in Fig.~\ref{fig:fig4}(c). When $\tau_2$ is small, the dominant error is single-photon loss, and ED itself can improve the fidelity significantly. However, as $\tau_2$ increases, more two-photon-loss errors accumulate, and the effectiveness of ED gradually diminishes. Then, combining ED with EP can further enhance the fidelity of the entangled state. Notably, the existence time of entanglement, corresponding to $\mathcal{N}>0$, increases by nearly $50\%$, extending from approximately $120$ to $180\,\mathrm{\mu s}$.}

	\textit{Conclusion--} We have experimentally demonstrated the first realization of entanglement pumping on bosonic logical qubits encoded in superconducting cavities. By combining quantum error detection with nonlocal entanglement purification techniques, we provide a resource-efficient strategy for preserving and enhancing the fidelity of entangled states between indirectly connected nodes in a quantum network. By reusing a pair of communication qubits and a pair of long-coherence storage cavities, we have demonstrated enhanced entanglement despite practical experimental imperfections, achieving a $50\%$ increase in {the existence time of entanglement within logical qubits}. {We also note the connections between entanglement pumping and dual-rail code, where entanglement pumping naturally implements erasure error detection.}
	Our results also highlight the potential of bosonic codes~\cite{cai2020bosonic,Joshi_2021} in quantum networks and provide new insights into entanglement management in modular quantum systems and distributed quantum information processing~\cite{Cirac1999,Cacciapuoti2020,PhysRevA.76.062323,PhysRevA.89.022317,PRXQuantum.2.017002}.
	
	\begin{acknowledgments}
		This work was funded by the Quantum Science and Technology-National Science and Technology Major Project (Grants No.~2024ZD0301500, No.~2021ZD0300200, and No.~2021ZD0301800) and the National Natural Science Foundation of China (Grants No. 92365301, No.~92565301, No.~12550006, No.~92265210, No.~92365206, No.~12474498, No.~12404567, No.~92165209). This work was also supported by the Fundamental Research Funds for the Central Universities and USTC Research Funds of the Double First-Class Initiative. This work was partially carried out at the USTC Center for Micro and Nanoscale Research and Fabrication.	
	\end{acknowledgments}
	
\bibliography{reference_proof}

\begin{thebibliography}{14}%
		\makeatletter
		\providecommand \@ifxundefined [1]{%
			\@ifx{#1\undefined}
		}%
		\providecommand \@ifnum [1]{%
			\ifnum #1\expandafter \@firstoftwo
			\else \expandafter \@secondoftwo
			\fi
		}%
		\providecommand \@ifx [1]{%
			\ifx #1\expandafter \@firstoftwo
			\else \expandafter \@secondoftwo
			\fi
		}%
		\providecommand \natexlab [1]{#1}%
		\providecommand \enquote  [1]{``#1''}%
		\providecommand \bibnamefont  [1]{#1}%
		\providecommand \bibfnamefont [1]{#1}%
		\providecommand \citenamefont [1]{#1}%
		\providecommand \href@noop [0]{\@secondoftwo}%
		\providecommand \href [0]{\begingroup \@sanitize@url \@href}%
		\providecommand \@href[1]{\@@startlink{#1}\@@href}%
		\providecommand \@@href[1]{\endgroup#1\@@endlink}%
		\providecommand \@sanitize@url [0]{\catcode `\\12\catcode `\$12\catcode
			`\&12\catcode `\#12\catcode `\^12\catcode `\_12\catcode `\%12\relax}%
		\providecommand \@@startlink[1]{}%
		\providecommand \@@endlink[0]{}%
		\providecommand \url  [0]{\begingroup\@sanitize@url \@url }%
		\providecommand \@url [1]{\endgroup\@href {#1}{\urlprefix }}%
		\providecommand \urlprefix  [0]{URL }%
		\providecommand \Eprint [0]{\href }%
		\providecommand \doibase [0]{http://dx.doi.org/}%
		\providecommand \selectlanguage [0]{\@gobble}%
		\providecommand \bibinfo  [0]{\@secondoftwo}%
		\providecommand \bibfield  [0]{\@secondoftwo}%
		\providecommand \translation [1]{[#1]}%
		\providecommand \BibitemOpen [0]{}%
		\providecommand \bibitemStop [0]{}%
		\providecommand \bibitemNoStop [0]{.\EOS\space}%
		\providecommand \EOS [0]{\spacefactor3000\relax}%
		\providecommand \BibitemShut  [1]{\csname bibitem#1\endcsname}%
		\let\auto@bib@innerbib\@empty
		\bibitem [{\citenamefont {Cai}\ \emph {et~al.}(2024)\citenamefont {Cai},
			\citenamefont {Mu}, \citenamefont {Wang}, \citenamefont {Zhou}, \citenamefont
			{Ma}, \citenamefont {Pan}, \citenamefont {Hua}, \citenamefont {Liu},
			\citenamefont {Xue}, \citenamefont {Yu}, \citenamefont {Wang}, \citenamefont
			{Song}, \citenamefont {Zou},\ and\ \citenamefont {Sun}}]{cai2024_S}%
		\BibitemOpen
		\bibfield  {author} {\bibinfo {author} {\bibfnamefont {W.}~\bibnamefont
				{Cai}}, \bibinfo {author} {\bibfnamefont {X.}~\bibnamefont {Mu}}, \bibinfo
			{author} {\bibfnamefont {W.}~\bibnamefont {Wang}}, \bibinfo {author}
			{\bibfnamefont {J.}~\bibnamefont {Zhou}}, \bibinfo {author} {\bibfnamefont
				{Y.}~\bibnamefont {Ma}}, \bibinfo {author} {\bibfnamefont {X.}~\bibnamefont
				{Pan}}, \bibinfo {author} {\bibfnamefont {Z.}~\bibnamefont {Hua}}, \bibinfo
			{author} {\bibfnamefont {X.}~\bibnamefont {Liu}}, \bibinfo {author}
			{\bibfnamefont {G.}~\bibnamefont {Xue}}, \bibinfo {author} {\bibfnamefont
				{H.}~\bibnamefont {Yu}}, \bibinfo {author} {\bibfnamefont {H.}~\bibnamefont
				{Wang}}, \bibinfo {author} {\bibfnamefont {Y.}~\bibnamefont {Song}}, \bibinfo
			{author} {\bibfnamefont {C.-L.}\ \bibnamefont {Zou}}, \ and\ \bibinfo
			{author} {\bibfnamefont {L.}~\bibnamefont {Sun}},\ }\href {\doibase
			10.1038/s41567-024-02446-8} {\bibfield  {journal} {\bibinfo  {journal} {Nat.
					Phys.}\ }\textbf {\bibinfo {volume} {20}},\ \bibinfo {pages} {1022} (\bibinfo
			{year} {2024})}\BibitemShut {NoStop}%
		\bibitem [{\citenamefont {Roy}\ \emph {et~al.}(2015)\citenamefont {Roy},
			\citenamefont {Kundu}, \citenamefont {Chand}, \citenamefont {Vadiraj},
			\citenamefont {Ranadive}, \citenamefont {Nehra}, \citenamefont {Patankar},
			\citenamefont {Aumentado}, \citenamefont {Clerk},\ and\ \citenamefont
			{Vijay}}]{RN9_S}%
		\BibitemOpen
		\bibfield  {author} {\bibinfo {author} {\bibfnamefont {T.}~\bibnamefont
				{Roy}}, \bibinfo {author} {\bibfnamefont {S.}~\bibnamefont {Kundu}}, \bibinfo
			{author} {\bibfnamefont {M.}~\bibnamefont {Chand}}, \bibinfo {author}
			{\bibfnamefont {A.~M.}\ \bibnamefont {Vadiraj}}, \bibinfo {author}
			{\bibfnamefont {A.}~\bibnamefont {Ranadive}}, \bibinfo {author}
			{\bibfnamefont {N.}~\bibnamefont {Nehra}}, \bibinfo {author} {\bibfnamefont
				{M.~P.}\ \bibnamefont {Patankar}}, \bibinfo {author} {\bibfnamefont
				{J.}~\bibnamefont {Aumentado}}, \bibinfo {author} {\bibfnamefont {A.~A.}\
				\bibnamefont {Clerk}}, \ and\ \bibinfo {author} {\bibfnamefont
				{R.}~\bibnamefont {Vijay}},\ }\href {\doibase 10.1063/1.4939148} {\bibfield
			{journal} {\bibinfo  {journal} {Appl. Phys. Lett.}\ }\textbf {\bibinfo
				{volume} {107}},\ \bibinfo {pages} {262601} (\bibinfo {year}
			{2015})}\BibitemShut {NoStop}%
		\bibitem [{\citenamefont {Kamal}\ \emph {et~al.}(2009)\citenamefont {Kamal},
			\citenamefont {Marblestone},\ and\ \citenamefont
			{Devoret}}]{PhysRevB.79.184301_S}%
		\BibitemOpen
		\bibfield  {author} {\bibinfo {author} {\bibfnamefont {A.}~\bibnamefont
				{Kamal}}, \bibinfo {author} {\bibfnamefont {A.}~\bibnamefont {Marblestone}},
			\ and\ \bibinfo {author} {\bibfnamefont {M.}~\bibnamefont {Devoret}},\ }\href
		{\doibase 10.1103/PhysRevB.79.184301} {\bibfield  {journal} {\bibinfo
				{journal} {Phys. Rev. B}\ }\textbf {\bibinfo {volume} {79}},\ \bibinfo
			{pages} {184301} (\bibinfo {year} {2009})}\BibitemShut {NoStop}%
		\bibitem [{\citenamefont {D\"ur}\ and\ \citenamefont
			{Briegel}(2007)}]{Dur_2007_S}%
		\BibitemOpen
		\bibfield  {author} {\bibinfo {author} {\bibfnamefont {W.}~\bibnamefont
				{D\"ur}}\ and\ \bibinfo {author} {\bibfnamefont {H.~J.}\ \bibnamefont
				{Briegel}},\ }\href {\doibase 10.1088/0034-4885/70/8/R03} {\bibfield
			{journal} {\bibinfo  {journal} {Rep. Prog. Phys.}\ }\textbf {\bibinfo
				{volume} {70}},\ \bibinfo {pages} {1381} (\bibinfo {year}
			{2007})}\BibitemShut {NoStop}%
		\bibitem [{\citenamefont {Werner}(1989)}]{PhysRevA.40.4277_S}%
		\BibitemOpen
		\bibfield  {author} {\bibinfo {author} {\bibfnamefont {R.~F.}\ \bibnamefont
				{Werner}},\ }\href {\doibase 10.1103/PhysRevA.40.4277} {\bibfield  {journal}
			{\bibinfo  {journal} {Phys. Rev. A}\ }\textbf {\bibinfo {volume} {40}},\
			\bibinfo {pages} {4277} (\bibinfo {year} {1989})}\BibitemShut {NoStop}%
		\bibitem [{\citenamefont {Deutsch}\ \emph {et~al.}(1996)\citenamefont
			{Deutsch}, \citenamefont {Ekert}, \citenamefont {Jozsa}, \citenamefont
			{Macchiavello}, \citenamefont {Popescu},\ and\ \citenamefont
			{Sanpera}}]{PhysRevLett.77.2818_S}%
		\BibitemOpen
		\bibfield  {author} {\bibinfo {author} {\bibfnamefont {D.}~\bibnamefont
				{Deutsch}}, \bibinfo {author} {\bibfnamefont {A.}~\bibnamefont {Ekert}},
			\bibinfo {author} {\bibfnamefont {R.}~\bibnamefont {Jozsa}}, \bibinfo
			{author} {\bibfnamefont {C.}~\bibnamefont {Macchiavello}}, \bibinfo {author}
			{\bibfnamefont {S.}~\bibnamefont {Popescu}}, \ and\ \bibinfo {author}
			{\bibfnamefont {A.}~\bibnamefont {Sanpera}},\ }\href {\doibase
			10.1103/PhysRevLett.77.2818} {\bibfield  {journal} {\bibinfo  {journal}
				{Phys. Rev. Lett.}\ }\textbf {\bibinfo {volume} {77}},\ \bibinfo {pages}
			{2818} (\bibinfo {year} {1996})}\BibitemShut {NoStop}%
		\bibitem [{\citenamefont {Khaneja}\ \emph {et~al.}(2005)\citenamefont
			{Khaneja}, \citenamefont {Reiss}, \citenamefont {Kehlet}, \citenamefont
			{Schulte-Herbrüggen},\ and\ \citenamefont {Glaser}}]{Khaneja2005_S}%
		\BibitemOpen
		\bibfield  {author} {\bibinfo {author} {\bibfnamefont {N.}~\bibnamefont
				{Khaneja}}, \bibinfo {author} {\bibfnamefont {T.}~\bibnamefont {Reiss}},
			\bibinfo {author} {\bibfnamefont {C.}~\bibnamefont {Kehlet}}, \bibinfo
			{author} {\bibfnamefont {T.}~\bibnamefont {Schulte-Herbrüggen}}, \ and\
			\bibinfo {author} {\bibfnamefont {S.~J.}\ \bibnamefont {Glaser}},\ }\href
		{\doibase https://doi.org/10.1016/j.jmr.2004.11.004} {\bibfield  {journal}
			{\bibinfo  {journal} {J. Magn. Reson.}\ }\textbf {\bibinfo {volume} {172}},\
			\bibinfo {pages} {296} (\bibinfo {year} {2005})}\BibitemShut {NoStop}%
		\bibitem [{\citenamefont {{de Fouquieres}}\ \emph {et~al.}(2011)\citenamefont
			{{de Fouquieres}}, \citenamefont {Schirmer}, \citenamefont {Glaser},\ and\
			\citenamefont {Kuprov}}]{deFouquieres2011_S}%
		\BibitemOpen
		\bibfield  {author} {\bibinfo {author} {\bibfnamefont {P.}~\bibnamefont {{de
						Fouquieres}}}, \bibinfo {author} {\bibfnamefont {S.}~\bibnamefont
				{Schirmer}}, \bibinfo {author} {\bibfnamefont {S.}~\bibnamefont {Glaser}}, \
			and\ \bibinfo {author} {\bibfnamefont {I.}~\bibnamefont {Kuprov}},\ }\href
		{\doibase https://doi.org/10.1016/j.jmr.2011.07.023} {\bibfield  {journal}
			{\bibinfo  {journal} {J. Magn. Reson.}\ }\textbf {\bibinfo {volume} {212}},\
			\bibinfo {pages} {412} (\bibinfo {year} {2011})}\BibitemShut {NoStop}%
		\bibitem [{\citenamefont {Johansson}\ \emph {et~al.}(2013)\citenamefont
			{Johansson}, \citenamefont {Nation},\ and\ \citenamefont {Nori}}]{RN112_S}%
		\BibitemOpen
		\bibfield  {author} {\bibinfo {author} {\bibfnamefont {J.~R.}\ \bibnamefont
				{Johansson}}, \bibinfo {author} {\bibfnamefont {P.~D.}\ \bibnamefont
				{Nation}}, \ and\ \bibinfo {author} {\bibfnamefont {F.}~\bibnamefont
				{Nori}},\ }\href {\doibase https://doi.org/10.1016/j.cpc.2012.11.019}
		{\bibfield  {journal} {\bibinfo  {journal} {Comput. Phys. Commun.}\ }\textbf
			{\bibinfo {volume} {184}},\ \bibinfo {pages} {1234} (\bibinfo {year}
			{2013})}\BibitemShut {NoStop}%
		\bibitem [{\citenamefont {Bland}\ \emph {et~al.}(2025)\citenamefont {Bland},
			\citenamefont {Bahrami}, \citenamefont {Martinez}, \citenamefont
			{Prestegaard}, \citenamefont {Smitham}, \citenamefont {Joshi}, \citenamefont
			{Hedrick}, \citenamefont {Kumar}, \citenamefont {Yang}, \citenamefont
			{Pakpour-Tabrizi}, \citenamefont {Jindal}, \citenamefont {Chang},
			\citenamefont {Cheng}, \citenamefont {Yao}, \citenamefont {Cava},
			\citenamefont {de~Leon},\ and\ \citenamefont {Houck}}]{RN5_S}%
		\BibitemOpen
		\bibfield  {author} {\bibinfo {author} {\bibfnamefont {M.~P.}\ \bibnamefont
				{Bland}}, \bibinfo {author} {\bibfnamefont {F.}~\bibnamefont {Bahrami}},
			\bibinfo {author} {\bibfnamefont {J.~G.~C.}\ \bibnamefont {Martinez}},
			\bibinfo {author} {\bibfnamefont {P.~H.}\ \bibnamefont {Prestegaard}},
			\bibinfo {author} {\bibfnamefont {B.~M.}\ \bibnamefont {Smitham}}, \bibinfo
			{author} {\bibfnamefont {A.}~\bibnamefont {Joshi}}, \bibinfo {author}
			{\bibfnamefont {E.}~\bibnamefont {Hedrick}}, \bibinfo {author} {\bibfnamefont
				{S.}~\bibnamefont {Kumar}}, \bibinfo {author} {\bibfnamefont
				{A.}~\bibnamefont {Yang}}, \bibinfo {author} {\bibfnamefont {A.~C.}\
				\bibnamefont {Pakpour-Tabrizi}}, \bibinfo {author} {\bibfnamefont
				{A.}~\bibnamefont {Jindal}}, \bibinfo {author} {\bibfnamefont {R.~D.}\
				\bibnamefont {Chang}}, \bibinfo {author} {\bibfnamefont {G.}~\bibnamefont
				{Cheng}}, \bibinfo {author} {\bibfnamefont {N.}~\bibnamefont {Yao}}, \bibinfo
			{author} {\bibfnamefont {R.~J.}\ \bibnamefont {Cava}}, \bibinfo {author}
			{\bibfnamefont {N.~P.}\ \bibnamefont {de~Leon}}, \ and\ \bibinfo {author}
			{\bibfnamefont {A.~A.}\ \bibnamefont {Houck}},\ }\href {\doibase
			10.1038/s41586-025-09687-4} {\bibfield  {journal} {\bibinfo  {journal}
				{Nature}\ }\textbf {\bibinfo {volume} {647}},\ \bibinfo {pages} {343}
			(\bibinfo {year} {2025})}\BibitemShut {NoStop}%
		\bibitem [{\citenamefont {Milul}\ \emph {et~al.}(2023)\citenamefont {Milul},
			\citenamefont {Guttel}, \citenamefont {Goldblatt}, \citenamefont {Hazanov},
			\citenamefont {Joshi}, \citenamefont {Chausovsky}, \citenamefont {Kahn},
			\citenamefont {\ifmmode~\mbox{\c{C}}\else \c{C}\fi{}ifty\"urek},
			\citenamefont {Lafont},\ and\ \citenamefont
			{Rosenblum}}]{PRXQuantum.4.030336_S}%
		\BibitemOpen
		\bibfield  {author} {\bibinfo {author} {\bibfnamefont {O.}~\bibnamefont
				{Milul}}, \bibinfo {author} {\bibfnamefont {B.}~\bibnamefont {Guttel}},
			\bibinfo {author} {\bibfnamefont {U.}~\bibnamefont {Goldblatt}}, \bibinfo
			{author} {\bibfnamefont {S.}~\bibnamefont {Hazanov}}, \bibinfo {author}
			{\bibfnamefont {L.~M.}\ \bibnamefont {Joshi}}, \bibinfo {author}
			{\bibfnamefont {D.}~\bibnamefont {Chausovsky}}, \bibinfo {author}
			{\bibfnamefont {N.}~\bibnamefont {Kahn}}, \bibinfo {author} {\bibfnamefont
				{E.}~\bibnamefont {\ifmmode~\mbox{\c{C}}\else \c{C}\fi{}ifty\"urek}},
			\bibinfo {author} {\bibfnamefont {F.}~\bibnamefont {Lafont}}, \ and\ \bibinfo
			{author} {\bibfnamefont {S.}~\bibnamefont {Rosenblum}},\ }\href {\doibase
			10.1103/PRXQuantum.4.030336} {\bibfield  {journal} {\bibinfo  {journal} {PRX
					Quantum}\ }\textbf {\bibinfo {volume} {4}},\ \bibinfo {pages} {030336}
			(\bibinfo {year} {2023})}\BibitemShut {NoStop}%
		\bibitem [{\citenamefont {Dong}\ \emph {et~al.}(2015)\citenamefont {Dong},
			\citenamefont {Zhang}, \citenamefont {Zou}, \citenamefont {Zou},\ and\
			\citenamefont {Guo}}]{Dong2015_S}%
		\BibitemOpen
		\bibfield  {author} {\bibinfo {author} {\bibfnamefont {D.}~\bibnamefont
				{Dong}}, \bibinfo {author} {\bibfnamefont {Y.-L.}\ \bibnamefont {Zhang}},
			\bibinfo {author} {\bibfnamefont {C.-L.}\ \bibnamefont {Zou}}, \bibinfo
			{author} {\bibfnamefont {X.-B.}\ \bibnamefont {Zou}}, \ and\ \bibinfo
			{author} {\bibfnamefont {G.-C.}\ \bibnamefont {Guo}},\ }\href {\doibase
			10.1016/j.physleta.2015.07.020} {\bibfield  {journal} {\bibinfo  {journal}
				{Phys. Rev. A}\ }\textbf {\bibinfo {volume} {379}},\ \bibinfo {pages} {2291}
			(\bibinfo {year} {2015})}\BibitemShut {NoStop}%
		\bibitem [{\citenamefont {Puri}\ and\ \citenamefont
			{Blais}(2016)}]{PhysRevLett.116.180501_S}%
		\BibitemOpen
		\bibfield  {author} {\bibinfo {author} {\bibfnamefont {S.}~\bibnamefont
				{Puri}}\ and\ \bibinfo {author} {\bibfnamefont {A.}~\bibnamefont {Blais}},\
		}\href {\doibase 10.1103/PhysRevLett.116.180501} {\bibfield  {journal}
			{\bibinfo  {journal} {Phys. Rev. Lett.}\ }\textbf {\bibinfo {volume} {116}},\
			\bibinfo {pages} {180501} (\bibinfo {year} {2016})}\BibitemShut {NoStop}%
		\bibitem [{\citenamefont {Paik}\ \emph {et~al.}(2016)\citenamefont {Paik},
			\citenamefont {Mezzacapo}, \citenamefont {Sandberg}, \citenamefont {McClure},
			\citenamefont {Abdo}, \citenamefont {C\'orcoles}, \citenamefont {Dial},
			\citenamefont {Bogorin}, \citenamefont {Plourde}, \citenamefont {Steffen},
			\citenamefont {Cross}, \citenamefont {Gambetta},\ and\ \citenamefont
			{Chow}}]{PhysRevLett.117.250502_S}%
		\BibitemOpen
		\bibfield  {author} {\bibinfo {author} {\bibfnamefont {H.}~\bibnamefont
				{Paik}}, \bibinfo {author} {\bibfnamefont {A.}~\bibnamefont {Mezzacapo}},
			\bibinfo {author} {\bibfnamefont {M.}~\bibnamefont {Sandberg}}, \bibinfo
			{author} {\bibfnamefont {D.~T.}\ \bibnamefont {McClure}}, \bibinfo {author}
			{\bibfnamefont {B.}~\bibnamefont {Abdo}}, \bibinfo {author} {\bibfnamefont
				{A.~D.}\ \bibnamefont {C\'orcoles}}, \bibinfo {author} {\bibfnamefont
				{O.}~\bibnamefont {Dial}}, \bibinfo {author} {\bibfnamefont {D.~F.}\
				\bibnamefont {Bogorin}}, \bibinfo {author} {\bibfnamefont {B.~L.~T.}\
				\bibnamefont {Plourde}}, \bibinfo {author} {\bibfnamefont {M.}~\bibnamefont
				{Steffen}}, \bibinfo {author} {\bibfnamefont {A.~W.}\ \bibnamefont {Cross}},
			\bibinfo {author} {\bibfnamefont {J.~M.}\ \bibnamefont {Gambetta}}, \ and\
			\bibinfo {author} {\bibfnamefont {J.~M.}\ \bibnamefont {Chow}},\ }\href
		{\doibase 10.1103/PhysRevLett.117.250502} {\bibfield  {journal} {\bibinfo
				{journal} {Phys. Rev. Lett.}\ }\textbf {\bibinfo {volume} {117}},\ \bibinfo
			{pages} {250502} (\bibinfo {year} {2016})}\BibitemShut {NoStop}%
	\end{thebibliography}
\bibliographystyle{Zou_proof}

	\global\long\def\figurename{ {Supplementary Figure}}%
	\renewcommand{\thefigure}{S\arabic{figure}}
	\setcounter{figure}{0}
	\global\long\def\thepage{S\arabic{page}}%
	\setcounter{page}{1}
	\global\long\def\theequation{S\arabic{equation}}%
	\setcounter{equation}{0}
	\global\long\def\tablename{\textbf{Supplementary Table}}%
	
	\begin{center}
		\textbf{{Supplemental Material}}
		\par\end{center}	
	\section{Experiment Setup and System Parameters}

	Our experiment utilizes two high-Q superconducting cavities ($S_1$ and $S_3$) to store entangled quantum states, each coupled to an $I$-type transmon ($I_1$ and $I_2$, respectively) that serves as a control qubit for manipulation. These two cavities are interconnected via two $Y$-type transmons ($Y_1$ and $Y_2$) functioning as communication qubits, along with another superconducting cavity ($S_2$). {The device used in this work is the same one as that in a previous work~\cite{cai2024_S}}.
	
\begin{figure*}
		\centering
		\includegraphics[scale=1]{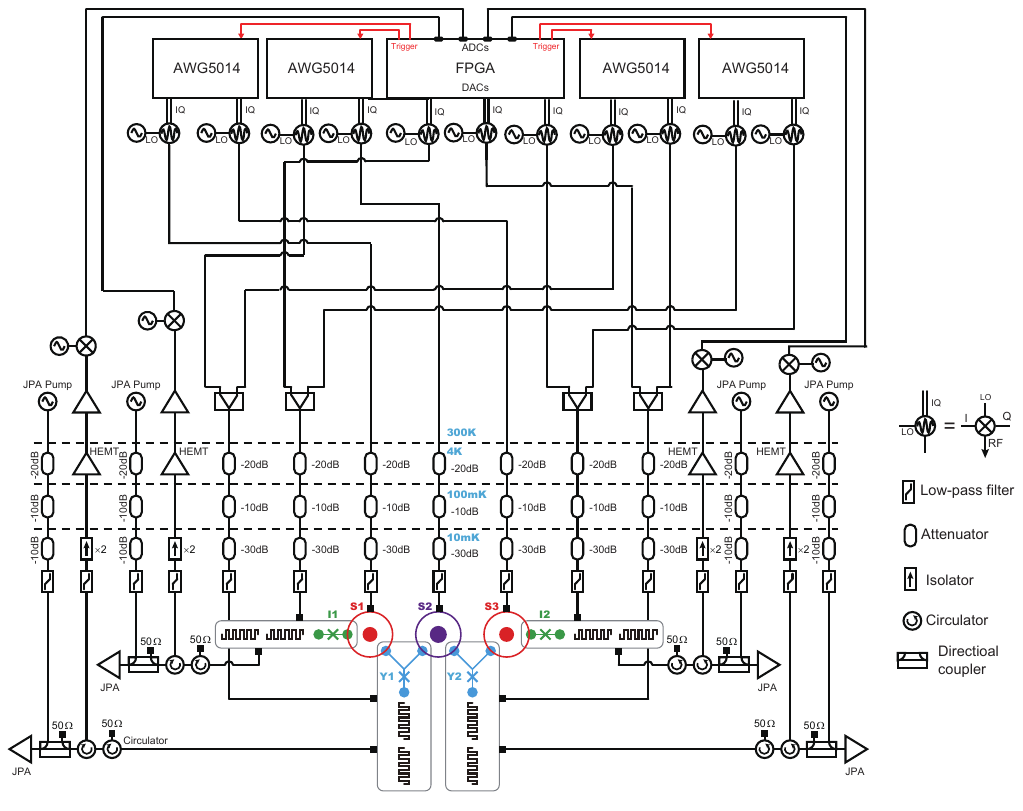}
		\caption[The system wiring]{\textbf{The system wiring.} 
			All input control signals are generated through an IQ mixing process. All the microwave lines are appropriately attenuated and filtered at different stages.
		}
		\label{fig:Purification_Wiring}
	\end{figure*}

	The detailed system wiring is shown in Fig.~\ref{fig:Purification_Wiring}. The device is placed on the base plate of a dilution refrigerator with a temperature of approximately 10~mK. All modes mentioned above are controlled by signals generated through an IQ mixing process, with the IQ signals produced by either arbitrary waveform generators (AWG) or the digital-to-analog converter (DAC) ports of field programmable gate arrays (FPGA). The output signals are amplified by Josephson parametric amplifiers (JPA)~\cite{RN9_S,PhysRevB.79.184301_S}, together with high-electron mobility transistors at 4 K.

	The dynamics of our system can be described by the  Hamiltonian
	\begin{equation} 
		\begin{aligned}
			H_{} =&\sum\limits_{i}K_i\hat a^\dagger_i\hat a^\dagger_i\hat a_i\hat a_i+\sum\limits_{i\neq j}\chi_{ij}\hat a^\dagger_i\hat a_i\hat a^\dagger_j\hat a_j,\\						
		\end{aligned}
		\label{eq:initial Hamiltonian}
	\end{equation}
	with $i,j \in \{S_1, S_2, S_3, I_1, I_2, Y_1, Y_2\}$. Here $K_i$ is the strength of the self-Kerr term of mode $i$, $\chi_{ij}$ is the cross-Kerr coupling strength between modes $i$ and $j$, and $\hat a_i$ is the annihilation operator of mode $i$. 
	Table~\ref{tab:SM_Hamiltonian} shows the measured nonlinear coupling strength in Eq.~\ref{eq:initial Hamiltonian}.  
	\begin{table}[h]
		\centering
		\caption{The measured nonlinear interactions ($\chi_{ij}/2 \pi $ in MHz) of the system (--- means the term is not measurable; $ \sim $ means this term is not measured).}
		\label{tab:SM_Hamiltonian}
		\resizebox{\linewidth}{!}{
			\begin{tabular}{c|ccccccc} 
				\hline
				Mode   & $I_1$  & $S_1$    & $Y_1$  & $S_2$    & $Y_2$             & $S_3$    & $I_2$      \\ 
				\hline
				$I_1$  & $\sim$ & 1.329    & 0.003  & ---      & 0.005             & ---      & ---        \\
				$S_1$  & 1.329  & 0.003    & 0.655  & $<0.001$ & ---               & ---      & ---        \\
				$Y_1$  & 0.003  & 0.655    & $\sim$ & 1.433    & 0.023             & ---      & ---        \\
				$S_2$  & ---    & $<0.001$ & 1.433  & 0.004    & 0.711             & $<0.001$ & ---        \\
				$Y_2$  & 0.005  & ---      & 0.023  & 0.711    & $\sim$            & 0.475    & 0.027      \\
				$S_3$  & ---    & ---      & ---    & $<0.001$ & 0.475             & 0.002    & 1.411      \\
				$I_2$  & ---    & ---      & ---    & ---      & 0.027             & 1.411    & $\sim$     \\
				\hline
			\end{tabular}
		}
	\end{table}
	
	\begin{table}[h]
		\centering
		\caption{The measured resonant mode frequencies, coherence times, and thermal populations.}
		\label{tab:SM_coherence}
		\begin{tabular}{c|cccc} 
			\hline
			Mode          & Frequency (GHz) & ${T_1}{\text{ (}}\mu {\text{s)}}$ & $T_2^*{\text{ (}}\mu {\text{s)}}$ & ${n_{th}}$ \\ 
			\hline
			$I_1$         & 4.232           & 80-110                               & 30-70                             & 3.5$\%$       \\
			$S_1$         & 6.102           & 261                               & ---                               & 3$\%$ 			\\
			$Y_1$         & 5.045           & 50-80                                & 20-40                                & 1$\%$       \\
			$S_2$         & 5.561           & 258                               & ---                               & 2$\%$      \\
			$Y_2$         & 4.707           & 80-120                           & 70-100                               & 3$\%$  \\
			$S_3$         & 6.006           & 256                               & ---                               & 2.5$\%$      \\
			$I_2$         & 4.479           & 65-85                            & 40-70                             & 4$\%$   \\
			\hline
		\end{tabular}
	\end{table}
	The coherence times and thermal populations for each mode are also measured in experiment, and the results are shown in Table ~\ref{tab:SM_coherence}. {The readout fidelities of the four transmon qubits are shown in Table ~\ref{tab:SM_readout_Fidelity}.}
	
	\begin{table}
		\caption{The readout fidelities of the transmon qubits.}
		\begin{tabular}{c|@{\hspace{10pt}} c @{\hspace{20pt}} c}
			\hline
			\centering
			& ${\rm F_{g}}$ &	${\rm F_{e}}$ \\ \hline
			$I_1$ & 0.9951 & 0.9922  \\
			$I_2$ & 0.9967 & 0.9878  \\
			$Y_1$ & 0.9982 & 0.9923 \\
			$Y_2$ & 0.9902 & 0.9848 \\
			\hline
		\end{tabular} \vspace{8pt}\\
		\label{tab:SM_readout_Fidelity}
	\end{table}
	
	\section{Entanglement pumping}
	\subsection{Principle}
	The details of the protocol of entanglement pumping (EP) can be found in Reference~\cite{Dur_2007_S}. For completeness, we explain the key principles in this section, starting with the basic entanglement purification protocol. Suppose we have two entangled pairs, with fidelities $F_A$ and $F_B$, respectively. For simplicity, we assume that the states are depolarized Werner state~\cite{PhysRevA.40.4277_S}, meaning that their density matrices can be written as:
	\begin{equation} 
		\begin{aligned}
			\rho_A=&F_A\ket{\psi_+}\bra{\psi_+}+\frac{1-F_A}{3}(\ket{\psi_-}\bra{\psi_-}+\ket{\phi_+}\bra{\phi_+}+\ket{\phi_-}\bra{\phi_-}),\\	
			\rho_B=&F_B\ket{\psi_+}\bra{\psi_+}+\frac{1-F_B}{3}(\ket{\psi_-}\bra{\psi_-}+\ket{\phi_+}\bra{\phi_+}+\ket{\phi_-}\bra{\phi_-}),\\					
		\end{aligned}
		\label{eq:rou}
	\end{equation}
	where $\ket{\psi_+}=\frac{\ket{01}+\ket{10}}{\sqrt{2}}$, $\ket{\psi_-}=\frac{\ket{01}-\ket{10}}{\sqrt{2}}$, $\ket{\phi_+}=\frac{\ket{00}+\ket{11}}{\sqrt{2}}$, and $\ket{\phi_-}=\frac{\ket{00}-\ket{11}}{\sqrt{2}}$. Here, $\ket{0}$ and $\ket{1}$ represent the two basis states of the qubits. If we want to increase the fidelity of the entangled pair labeled as A, we can use the two qubits in party A as control qubits and apply CNOT operations to the two qubits in party B, as shown in Fig.~\ref{fig:EPsimu}(a). 
	
	Afterward, we measure the two qubits in party B and retain the quantum states in party A if the measurement results are the same. The fidelity of the entangled pair that is retained in party A will then become
	\begin{equation} 
		\begin{aligned}
			F'=\frac{F_AF_B+(\frac{1-F_A}{3})(\frac{1-F_B}{3})}{F_AF_B+F_A(\frac{1-F_B}{3})+(\frac{1-F_A}{3})F_B+5(\frac{1-F_A}{3})(\frac{1-F_B}{3})}.\\						
		\end{aligned}
		\label{eq:fidelity}
	\end{equation}
	
	{The success probability of obtaining identical measurement results becomes}
	{\begin{equation} 
			\begin{aligned}
				p_\mathrm{success}=\frac{5+8F_AF_B-2(F_A+F_B)}{9}.					
			\end{aligned}
			\label{eq:successrate}
	\end{equation}}
	{In the experiment, we post-select the measurement results where both communication qubits are at in the ground state $\ket{gg}$ to avoid resetting them. The success probability of selecting $\ket{gg}$ satisfies $p_{\ket{gg}}=\frac{p_\mathrm{success}}{2} $. }
	
	For a given $F_B$, there exists an upper limit $F_\mathrm{limit}$ for purification. If $F_A < F_\mathrm{limit}$, the fidelity of the entangled state in party A will increase after purification; if $F_A > F_\mathrm{limit}$, the fidelity will decrease after purification. Making $F'=F_A$, we can solve Eq.~\ref{eq:fidelity} and get the fidelity limit:
	\begin{equation} 
		\begin{aligned}
			F_\mathrm{limit}=\frac{3(2F_B-1)+\sqrt{28F_B^2-26F_B+7}}{2(4F_B-1)}.\\						
		\end{aligned}
		\label{eq:limitfidelity}
	\end{equation}
	
	We plot the curve of $F_\mathrm{limit}$ in Fig.~\ref{fig:EPsimu}(b). The re-entangled fidelity realized in our experiment (92.3\%) is indicated in the figure by a gray arrow, and the corresponding $F_\mathrm{limit}= 95.8\%$.
	
	The basic idea of EP is to repeatedly generate entangled states in party B and use them to purify the entangled state in party A over and over again. If any purification step fails, we need to start over the entire process. Although this method sacrifices the success probability, it can always push the fidelity of the entangled state from a smaller value to $F_\mathrm{limit}$. 
	
	However, it should be noted that during the EP process, the entangled state in party A will gradually deviates from the depolarized Werner state, and errors will accumulate in $\ket{\psi^-}\bra{\psi^-}$ subspace. This is caused by two reasons. First, the entanglement purification process increases not only the proportion of $\ket{\psi^+}\bra{\psi^+}$ but also the proportion of $\ket{\psi^-}\bra{\psi^-}$. Second, single-qubit dephasing errors may convert state $\ket{\psi^+}$ to $\ket{\psi^-}$. When the proportion of $\ket{\psi^-}\bra{\psi^-}$ is too large, purification will no longer improve, and may even reduce the fidelity of the entangled state. Therefore, we need to perform local operations to stop the accumulation of errors. In our experiment, we employ the DEJMPS protocol~\cite{PhysRevLett.77.2818_S}. After every purification step, we perform an $R_x(\frac{\pi}{2})$ transformation on cavity $S_1$ and an $R_x(-\frac{\pi}{2})$ transformation on cavity $S_3$. After the transformations, the proportions of $\ket{\psi^-}\bra{\psi^-}$ and $\ket{\phi^-}\bra{\phi^-}$ are swapped. As a result, the proportions of all three kinds of error states can be reduced continuously. 
	
	\subsection{Gates optimized by GRAPE}
	As discussed in the main text, in order to realize EP, we need several quantum gates including two-qubit ($Y_1$ and $Y_2$) entanglement and re-entanglement gates, encode-decode gates, CNOT gates between cavities and transmon qubits, and local operation gates on cavities. In our experiment, all these gates are realized via numerical optimization using the gradient ascent pulse engineering method (GRAPE)~\cite{Khaneja2005_S,deFouquieres2011_S}. The communication qubit entanglement gates and encode-decode gates have been introduced in previous works~\cite{cai2024_S}. In this section, we will mainly discuss the implementation of the re-entanglement gates, CNOT gates, and local operations.

	The goal of the re-entanglement gate is to regenerate entanglement on the communication qubits $Y_1$ and $Y_2$ under the condition that an entangled state already exists between the storage cavities $S_1$ and $S_3$. Since the coupling between the cavity and the qubit cannot be turned off, the entanglement gate optimized by considering only the modes $Y_1$, $Y_2$, and $S_2$  will no longer function properly. Therefore, when optimizing the re-entanglement gate, we need to take into account all five modes: $Y_1$, $Y_2$, $S_1$, $S_2$, and $S_3$. 
	
	For optimization, we define the initial state $\ket{\psi_{i}}$ and the target state $\ket{\psi_{t}}$ as constraints. For the regular entanglement gate, when both $S_1$ and $S_3$ are in their vacuum states, the constraints are given by
	\begin{equation} 
		\begin{aligned}
			\ket{\psi_{i}}&=(\ket{g}\ket{0}\ket{g})_{Y_1S_2Y_2},\\			\ket{\psi_{t}}&=\frac{1}{\sqrt2}(\ket{g}\ket{0}\ket{e}+\ket{e}\ket{0}\ket{g})_{Y_1S_2Y_2},\\			
		\end{aligned}
		\label{eq:entanglegate}
	\end{equation}
	where $\ket{g}$ and $\ket{e}$ represent the ground and excited states of the communication qubits, and $\ket{0}$ stands for the vacuum state of cavity $S_2$. For the re-entanglement gate, when $S_1$ and $S_3$ are encoded in $\{\ket{0},\ket{1}\}$, the constraints become 
	\begin{equation} 
		\begin{aligned}
			\ket{\psi_{i}}&=\frac{1}{\sqrt2}(\ket{0}\ket{1}+\ket{1}\ket{0})_{S_1S_3}\otimes(\ket{g}\ket{0}\ket{g})_{Y_1S_2Y_2},\\			\ket{\psi_{t}}&=\frac{1}{2}(\ket{0}\ket{1}+\ket{1}\ket{0})_{S_1S_3}\otimes(\ket{g}\ket{0}\ket{e}+\ket{e}\ket{0}\ket{g})_{Y_1S_2Y_2}.\\			
		\end{aligned}
		\label{eq:entanglegate2}
	\end{equation}
	It is worth noting that for both of the above gates, the drives are applied only to the modes $Y_1$, $S_2$, and $Y_2$. Therefore, we can truncate $S_1$ and $S_3$ to the first two levels. As a result, the dimension of the time evolution operators becomes 16 times the original size, which is acceptable in numerical computation. However, for binomial encoding, the truncation extends to the first five energy levels, and the computational dimension increases to 625 times the original size, which is unacceptable. Therefore, we employ the echoed cavity-induced phase (CIP) gate to realize the re-entanglement gate for the entangled logical qubits (ELQ) in the storage cavities, as described in the main text.
	
	For the CNOT gates and the local operations, both involve only two modes: $S_{1(3)}$ and $Y_{1(2)}$. Therefore, they can be relatively easily optimized. For simplicity, we have combined these two operations into a single unitary for optimization, which can be expressed as $U=U_\mathrm{local}U_\mathrm{CNOT}$.
	
	{Since we cannot directly measure the cavity, we also employ optimized pulses in the process of state tomography. To perform the state tomography, we first apply the decoding pulses to the storage cavity $S_{1(3)}$ and the corresponding control qubit $I_{1(2)}$. The conditions for optimization are:
		\begin{equation} 
			\begin{aligned}
				|0_L\rangle|g\rangle &\rightarrow |0\rangle|g\rangle, \\
				\frac{|0_L\rangle + |1_L\rangle}{\sqrt{2}}|g\rangle &\rightarrow |0\rangle\frac{|g\rangle + |e\rangle}{\sqrt{2}},			
			\end{aligned}
			\label{eq:decode}
		\end{equation}
		where for a physical qubit encoding $\ket{0_L}=\ket{0}$, $\ket{1_L}=\ket{1}$, and for a logical qubit encoding $\ket{0_L}=(\ket{0}+\ket{4})/\sqrt{2}$, $\ket{1_L}=\ket{2}$. After applying these pulses, the quantum information is decoded from cavities to control qubits. Then, we utilize single-qubit rotations and measurements to obtain the density matrix of the state from the control qubits.}
	\subsection{Numerical simulation}
	As described above, EP can increase the fidelity of the entangled state to $F_\mathrm{limit}$ with ideal operations. However, due to factors such as qubit decoherence and measurement errors, the actual converged fidelity will always be lower than $F_\mathrm{limit}$. To assess the impact of decoherence on the final fidelity, we carry out numerical simulations~\cite{RN112_S} of the EP process with the parameters in Tables~\ref{tab:SM_Hamiltonian} and \ref{tab:SM_coherence} and the experimentally measured density matrix of the re-entangled state. 
	
\begin{figure}
		\includegraphics{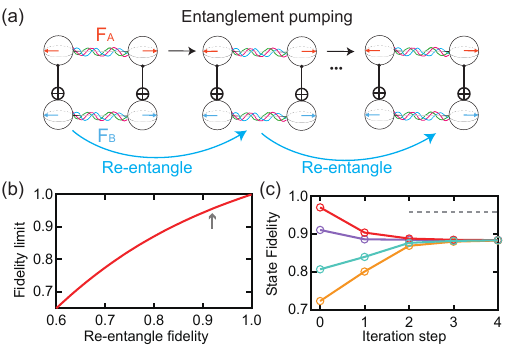}
		\caption{\textbf{Upper bound of the EP fidelity and simulation results.} (a) Schematic of the EP protocol. (b) Theoretical limit of the converged fidelity for the purified entangled state varies with the re-entangled state fidelity. The measured re-entangled state fidelity (92.3\%) is marked by the array. (c) Simulation results of the EP process for different initial states. We adopt the measured re-entangled state and include the decoherence of the communication qubits and cavities during the CNOT gates and local operations. The theoretical upper bound of the converged fidelity is indicated by the dashed line.}
		\label{fig:EPsimu}
	\end{figure}

	
	The simulation results are shown in Fig.~\ref{fig:EPsimu}(c). We choose four Werner states with different fidelities as the initial states. After three rounds of EP, their fidelities all converge to around 88.2\%, which is 7.6\% lower than the ideal fidelity limit (shown as dashed line). The data presented in the main text, however, converge to around 86.7\%, which is only slightly lower than the simulation results. This indicates that the final fidelity is primarily limited by decoherence. The difference between simulation and experiment could be attributed to errors not included in the simulation, such as measurement errors as shown in Table~\ref{tab:SM_readout_Fidelity} for post-selection.
	

	To further quantify the contributions of cavity decay and qubit decoherence to the final fidelity, we conduct additional analysis. Since we have performed the local operations, the final steady state can be approximately considered as a depolarized Werner state:
	\begin{equation} 
		\begin{aligned}
			\rho_f=&x\ket{\psi_+}\bra{\psi_+}+\frac{1-x}{4}\hat I,\\					
		\end{aligned}
		\label{eq:rou2}
	\end{equation}
	where x is the depolarizing ratio. It should be noted that Eq.~\ref{eq:rou} and Eq.~\ref{eq:rou2} are two equivalent representations, and the ratio $x$ and state fidelity $F$ satisfy $x=\frac{4F-1}{3}$. Assuming that the effects of various errors on the final fidelity are independent, the ratio x can be expressed as
	\begin{equation} 
		x=\prod \limits_{i} x_i,
		\label{eq:ratio}
	\end{equation}
	where $i$ represents different types of errors. To extract the ratio corresponding to each type of error, we add only the specific type of error in the simulation, obtain the final converged fidelity, and convert it into $x_i$.  
	
	\begin{table}
		\centering
		\caption{Error budgets of our experiment. $F_i$ and $x_i$ are the fidelity and depolarizing ratio, respectively, for the corresponding error type.}
		\label{tab:error buget}
		\begin{tabular}{c|@{\hspace{10pt}} c @{\hspace{20pt}} c} 
			\hline
			error type         & $F_i$ & $x_i$   \\ 
			\hline
			Re-entangle        & 0.961   & 0.948   \\
			Qubit decoherence  & 0.944   & 0.925   	\\
			Cavity decay       & 0.974   & 0.965     \\
			All error          & 0.882   & 0.843      \\
			experiment         & 0.867   & 0.823       \\
			\hline
		\end{tabular}
	\end{table}

	The simulation results are shown in Table~\ref{tab:error buget}. The data in the row labeled as ``re-entangle" is obtained from simulation with experimentally measured re-entangled density matrix and ideal operations. The resulting fidelity is very close to the theoretically predicted upper limit (0.958), and the difference is likely due to the simulation precision. From the data in Table~\ref{tab:error buget}, we can see that qubit decoherence contributes the most to the infidelity. If we multiply $x_i$ of the first three rows, we obtain $\prod \limits_i x_i=0.846$, which aligns to the one obtained from the simulation with all errors included (0.843). This justifies the assumption of independent errors.

	{Given that the duration of purification is much shorter than the system's coherence time, the converged fidelity is expected to exhibit a linear dependence on the system coherence. To quantify how converged fidelity depends on coherence parameters, we change the parameters $(T_1, T_{\phi}, T_c)$ independently in the simulation and perform linear fittings. We obtain the relationship 
		\begin{equation} 
			\begin{aligned}
				F=0.944(1-\frac{0.577 }{T_{1,Y_1}}) (1-\frac{0.496 }{T_{\phi,Y_1} })(1- \frac{3.694}{T_{1,S_1} })\\
				(1- \frac{0.671}{T_{1,Y_2}}) (1-\frac{0.635}{T_{\phi,Y_2}}) (1-\frac{3.969}{T_{1,S_3}}),
			\end{aligned}
			\label{eq:Gate error}
		\end{equation}
		where $T_{1,Y_{1(2)}}$ and $T_{\phi,Y_{1(2)}}$ are the corresponding communication qubit lifetime and dephasing time in units of $\mu$s, respectively, and  $T_{c,S_{1(3)}}$ is the lifetime of the storage cavity in units of $\mu$s. The coefficient $0.944$ is mainly limited by the fidelity of the re-entangled states and the decoherence of the control qubits $I_1$ and $I_2$ during the decoding process.  }

	{Since the entangled state is continuously stored in the two cavities, the EP process is particularly sensitive to their coherence times, as also indicated by the coefficients in Eq.~\ref{eq:Gate error}. To illustrate the impact of improved coherence times on the performance of EP, we simulate the entire process, with all cavity-qubit gate operations implemented using GRAPE (including the re-entanglement gate). In the simulation, the ground state readout fidelity is $99.9\%$, and the thermal populations of both the cavities and qubits are fixed at $1\%$. If the cavity lifetime is increased to 10~ms and the qubit relaxation and dephasing times reach $500~\mathrm{\mu}$s, all of which are state-of-the-art experimental values ~\cite{RN5_S,PRXQuantum.4.030336_S},  simulations show that the resulting saturated entanglement fidelity can reach $99.1\%$. In comparison, when the cavity lifetime is 1~ms, the saturated entanglement fidelity is reduced to $98.5\%$.}

	\begin{figure*}
		\centering
		\includegraphics{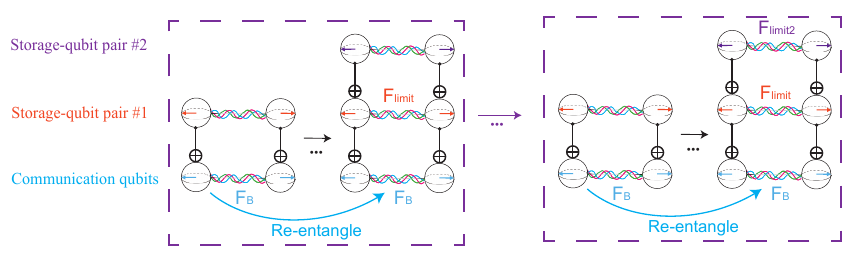}
		\caption[Multi-layer entanglement pumping]{\textbf{Schematic for two-layer EP.} 
		}
		\label{fig:EPML}
	\end{figure*}
	
	\subsection{Multi-layer entanglement pumping}
	
	{EP scheme is important because it provides a way to maximize the fidelity of remotely prepared entangled states under the constraints of limited hardware resources. Equation~\ref{eq:limitfidelity} provides the fidelity limit when there is only one pair of communication qubits and one pair of storage qubits. In fact, we can further reduce the infidelity by increasing the number of storage qubits and employing multi-layer EP.}
	
	Figure~\ref{fig:EPML} shows the schematic for two-layer EP. We first use the communication qubits to increase the fidelity of entangled storage-qubit pair \#1 to $F_\mathrm{limit}$, and then use storage-qubit pair \#1 to perform purification for storage-qubit pair \#2. By repeating this process, we can increase the entanglement fidelity of storage-qubit pair \#2 to $F_\mathrm{limit2}$, which can be calculated by replacing $F_{B}$ in Eq.~\ref{eq:limitfidelity} with $F_\mathrm{limit}$.This process can be generalized to an arbitrary number of storage-qubit pairs, thereby increasing the entanglement fidelity to an arbitrarily high level. {Specifically, to realize this architecture, each node consists of an ancillary qubit and a series of linearly connected cavities, with a transmon qubit placed between each pair of neighboring cavities to enable CNOT and SWAP operations. Initially, EP is repeatedly performed between the communication qubits and storage-qubit pair \#1 until saturation is reached. The entangled state is then swapped to storage-qubit pair \#2. Subsequently, the communication qubits continue to pump entanglement with storage-qubit pair \#1 until saturation, and this entangled state is used to further pump the entanglement of storage-qubit pair \#2 to a higher level. The state is then swapped to storage-qubit pair \#3, and the process is repeated iteratively. In this way, multi-layer EP is achieved.}
	
	{Assume that the infidelity of the initial entangled state, $\epsilon = 1 - F_B$, is a small quantity, i.e., $\epsilon \ll 1$. Then, using Eq.~\ref{eq:limitfidelity}, the infidelity after the first layer EP can be expressed as
		\begin{equation} 
			\begin{aligned}
				\epsilon_1 = 1 - F_{\text{limit}} = \epsilon/2 + O(\epsilon^2).
			\end{aligned}
			\label{eq:MPerror}
		\end{equation} 
		As a result, after $n$ layers of EP, the infidelity becomes $\epsilon_n = \epsilon / 2^n$. Therefore, by linearly increasing the number of storage-qubit pairs $n$, we can exponentially suppress the infidelity. However, it should be noted that the cost of this approach is an exponential decrease in the success rate of post-selection, a trade-off also inherent to traditional entanglement purification schemes.}
	
	\section{Entanglement pumping and parity difference measurement}
	\subsection{Parity difference measurement}
	{As discussed in the main text, replacing the CNOT gates in the entanglement purification protocol with parity mapping operations enables determination of whether the two cavities share identical parity. In this section, we will provide a detailed analysis.}
	
	{In the experiment, we use three types of CNOT-like gates: GRAPE, {mod-2 parity mapping}, and {mod-4 parity mapping}. The process matrices of the different gates are reconstructed through {simulation}, allowing to calculate the process fidelity using $F_\chi=\mathrm{Tr}(R_\mathrm{ideal}^\dagger R)/d^2$ as shown in Table~\ref{tab:CNOTgatePF}.}
	
	{The regular mod-2 parity mapping operation consists of two $\pi/2$ rotations separated by a free-evolution time $t = \pi/\chi$, where $\chi$ is the dispersive coupling strength between the communication qubit and the storage cavity. 
		During the free-evolution period, the qubit excited state $\ket{e}$ accumulates a phase of $\pi$ when the cavity is in odd states, while acquiring no extra phase for even cavity states. Thus, this parity mapping operation implements a parity-dependent gate on the communication qubit: an identity operation for an even parity while an X-gate for an odd parity.}
	
	{Consider the two communication qubits initially prepared in the entangled state $(\ket{ge}+\ket{eg})/\sqrt2$. After applying bilateral mod-2 parity mapping operations, the final state depends on the parities of the two cavities as follows:
		\begin{equation} 
			\begin{aligned}
				&\ket{\mathrm{even}}\ket{\mathrm{even}} \otimes \frac{1}{\sqrt2}(\ket{ge}+\ket{eg}),\\	
				&\ket{\mathrm{odd}}\ket{\mathrm{odd}} \otimes \frac{1}{\sqrt2}(\ket{ge}+\ket{eg}),\\	
				&\ket{\mathrm{even}}\ket{\mathrm{odd}} \otimes \frac{1}{\sqrt2}(\ket{gg}+\ket{ee}),\\	
				&\ket{\mathrm{odd}}\ket{\mathrm{even}} \otimes \frac{1}{\sqrt2}(\ket{gg}+\ket{ee}).\\					
			\end{aligned}
			\label{eq:paritydiff}
		\end{equation}
		Therefore, by measuring the communication qubits and determining whether their states are the same, we can directly determine whether the two cavities have identical parity. This effectively implements a measurement of the parity difference $e^{i\pi(\hat n_1-\hat n_2)}$, where $\hat n_{1(2)}$ are the photon numbers in the two storage cavities.}
	
	{Similarly, the mod-4 parity mapping operation also consists of two $\pi/2$ rotations, but separated by a free-evolution time $t = \pi/2\chi$. 
		During the free-evolution period, the qubit excited state $\ket{e}$ accumulates a phase of $\pi$ when the cavity photon number $n=2$ (mod 4), while acquiring no extra phase for $n=0$ (mod 4). 
		Therefore, this parity mapping operation implements a mod-4 parity-dependent gate on the communication qubit: an identity operation for $n=0$ (mod 4) while an X-gate for $n=2$ (mod 4).}	
	
	{\begin{table}
			\centering
			\caption{The simulated process fidelity of different CNOT-like gates.}
			\label{tab:CNOTgatePF}
			\begin{tabular}{c|@{\hspace{10pt}} c@{\hspace{10pt}} |@{\hspace{10pt}} c@{\hspace{10pt}} |@{\hspace{10pt}} c } 
				\hline
				Gate & System & Length & $F_{\chi}$ \\ \hline
				GRAPE & $Y_1\otimes S_1$ & $3~\mu s$ & 93.75\% \\
				Mod-2 parity mapping & $Y_1\otimes S_1$ & $0.784~\mu s$ & 98.33\% \\
				Mod-4 parity mapping & $Y_1\otimes S_1$ & $0.402~\mu s$ & 97.88\% \\
				GRAPE & $Y_2\otimes S_3$ & $3~\mu s$ & 97.02\% \\
				Mod-2 parity mapping & $Y_2\otimes S_3$ & $1.076~\mu s$ & 98.90\% \\
				Mod-4 parity mapping & $Y_2\otimes S_3$ & $0.548~\mu s$ & 98.51\% \\ \hline
			\end{tabular}
	\end{table}}

	\subsection{Concatenation of binomial logical qubits with dual-rail code.}
	
	{The dual-rail qubits encoded in the $\ket{0}$ and $\ket{1}$ states of the cavities can be replaced with error-correctable binomial logical qubits. The concatenated states can be expressed as $(\alpha\ket{0_L}\ket{1_L}+\beta\ket{1_L}\ket{0_L})$, where $\ket{0_L}=({\ket{0}+\ket{4}})/{\sqrt 2}$ and $\ket{1_L}=\ket{2}$.  {For the dual-rail code encoded in $\ket{0}$ and $\ket{1}$, an erasure error occurs will map $\ket{1}$ to $\ket{0}$. Since single-photon-loss errors are intrinsically protected by the binomial encoding, the erasure error in the dual-rail code encoded in $\ket{0_L}$ and $\ket{1_L}$ corresponds to a two-photon-loss error. In this case, both logical states are affected, with $\ket{0_L}$ mapped to $\ket{1_L}$ and $\ket{1_L}$ mapped to $\ket{0}$.}
		This concatenation can provide two-layer protections. First, with the binomial encoding we can perform error detection through parity measurement to mitigate the single-photon-loss errors. Second, by replacing the CNOT gates in the EP scheme with the mod-4 parity mapping operations, we can determine whether the two cavities share identical mod-4 parity, {provided that the communication qubits have been re-entangled to $ \frac{1}{\sqrt2}(\ket{ge}+\ket{eg})$}. If no error occurs, the mod-4 parity mapping operations yield the final state $ (\alpha\ket{0_L}\ket{1_L}+\beta\ket{1_L}\ket{0_L}) \otimes \frac{1}{\sqrt2}(\ket{gg}+\ket{ee})$, with the encoded quantum information unaffected. For the case where an erasure error occurs, the final state will become $ \ket{0_L}\ket{0_L} \otimes \frac{1}{\sqrt2}(\ket{ge}+\ket{eg})$. Without loss of generality, if a two-photon-loss error occurs in the first cavity, the final state will become {$ (\sqrt6\alpha\ket{2}\ket{1_L}+\beta\ket{0}\ket{0_L}) \otimes \frac{1}{\sqrt2}(\ket{ge}+\ket{eg})$}. Although the quantum information is affected, we can suppress the errors by discarding results where the communication qubits are found in different states.} 
	
	{It is noted that the binomial ELQ is actually a superposition state of the concatenated code. Therefore, the EP for the binomial qubits in the main text can be interpreted as a demonstration of the concatenation of binomial codes with dual-rail encoding.}

	\begin{table}[h]
		\centering
		\caption{Final states of the cavities and ancillary qubits after EP for various error types.}
		\label{tab:EPtheory}
		\begin{tabular}{|c|c|c|c|}
			\hline
			Error order & Error type & Cavity state (after error) & \makecell{Ancilla state\\(after EP)} \\
			\hline
			$E^{(0)}$& $I \otimes I$ & $\alpha \ket{0_L}\ket{1_L} + \beta \ket{1_L}\ket{0_L}$ & $\ket{\phi^+}$ \\
			\hline
			$E^{(1)}$&$I \otimes \hat{b}$ & $  \sqrt{2}\alpha \ket{0_L}\ket{1_E} + 2\beta \ket{1_L}\ket{0_E} $ & $\frac{\ket{\phi^+}-i\ket{\psi^+}}{\sqrt{2}}$ \\
			\hline
			$E^{(1)}$&$\hat{a} \otimes I$ & $  2\alpha \ket{0_E}\ket{1_L} + \sqrt{2}\beta \ket{1_E}\ket{0_L} $ & $\frac{\ket{\phi^+}-i\ket{\psi^+}}{\sqrt{2}}$ \\
			\hline
			$E^{(2)}$&$\hat{a} \otimes \hat{b}$ & $\alpha \ket{0_E}\ket{1_E} + \beta \ket{1_E}\ket{0_E}$ & $\ket{\psi^+}$ \\
			\hline
			$E^{(2)}$&$I \otimes \hat{b}^2$ & $  \alpha \ket{0_L}\ket{0} + \sqrt{6}\beta \ket{1_L}\ket{2} $ & $\ket{\psi^+}$ \\
			\hline
			$E^{(2)}$&$\hat{a}^2 \otimes I$ & $  \sqrt{6}\alpha \ket{0_L}\ket{0} + \beta \ket{1_L}\ket{2} $ & $\ket{\psi^+}$ \\
			\hline
			$E^{(3)}$&$\hat{a} \otimes \hat{b}^2$ & $  \alpha \ket{3}\ket{0} + \sqrt{3}\beta \ket{1}\ket{2} $ & $\frac{\ket{\phi^+}+i\ket{\psi^+}}{\sqrt{2}}$ \\
			\hline
			$E^{(3)}$&$\hat{a}^2 \otimes \hat{b}$ & $  \sqrt{3}\alpha \ket{3}\ket{0} + \beta \ket{1}\ket{2} $ & $\frac{\ket{\phi^+}+i\ket{\psi^+}}{\sqrt{2}}$ \\
			\hline
			$E^{(4)}$&$\hat{a}^2 \otimes \hat{b}^2$ & $\alpha \ket{2}\ket{0} + \beta \ket{0}\ket{2}$ & $\ket{\phi^+}$ \\
			\hline
		\end{tabular}
	\end{table}
	
	The capability of EP to mitigate different types of errors in the two cavities is analyzed in the following. The EP process essentially implements a mod-4 parity measurement, corresponding to a transformation $Y\Big(-\frac{\pi}{2}\Big) Z\Big(\frac{N}{4} \times 2\pi\Big) Y\Big(\frac{\pi}{2}\Big)$ on the ancillary qubit, where $N$ is the photon number in the cavity adjacent to the auxiliary qubit. We consider the initial state of the auxiliary qubits as $\ket{\psi^+}=\frac{1}{\sqrt{2}} (\ket{g}\ket{e} + \ket{e}\ket{g})$ and the initial state of the cavities as $\alpha \ket{0_L}\ket{1_L} + \beta \ket{1_L}\ket{0_L}$. The resulting changes in the final states for various cavity errors are summarized in Table~\ref{tab:EPtheory}, where $\ket{0_L}=\frac{1}{\sqrt2}(\ket{0}+\ket{4}),\ket{0_E}=\ket{3},\ket{1_L}=\ket{2},\ket{1_E}=\ket{1},\ket{\psi^+}=\frac{1}{\sqrt{2}} (\ket{g}\ket{e} + \ket{e}\ket{g}), \ket{\phi^+}=\frac{1}{\sqrt{2}} (\ket{g}\ket{g} + \ket{e}\ket{e})$. It is worth noting that after EP, the $\ket{0_L}$ state changes from $\frac{\ket{0}+\ket{4}}{\sqrt{2}}$ to $\frac{\ket{0}-\ket{4}}{\sqrt{2}}$, which requires a phase correction or an adjustment of the default coordinate axes.
	
	Therefore, by measuring whether the final ancilla states are identical, one can completely mitigate second-order errors including unilateral two-photon-loss errors ($I \otimes \hat{b}^2$ and $\hat{a}^2 \otimes I$) and bilateral single-photon-loss errors ($\hat{a} \otimes \hat{b}$). EP can also mitigate $50\%$ of first-order errors ($\hat{a}\otimes I$ and $I\otimes \hat{b}$) and third-order errors ($\hat{a}^2\otimes \hat{b}$ and $\hat{a}\otimes \hat{b}^2$). However, it does not provide any suppression for bilateral two-photon-loss errors ($\hat{a}^2\otimes \hat{b}^2$). Subsequent error detection (ED), performed via mod-2 parity measurements, can fully mitigate any single-photon loss in either cavity. Consequently, by combining ED with EP, errors in the two-cavity system can be suppressed down to only bilateral two-photon-loss errors.

	\section{Echoed Cavity-induced Phase gate}
	In the main text, we briefly discuss the basic idea of echoed CIP gate. In this section, we will provide more information about the principle of CIP gate and the calibration experiment.
	
	\subsection{Principle of CIP gate}
\begin{figure}
		\includegraphics{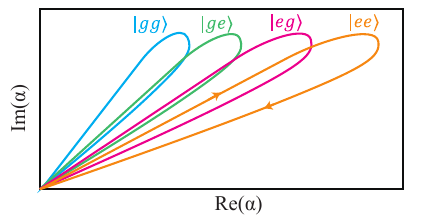}
		\caption{\textbf{Diagram for the CIP gate.} The state of the storage cavity adiabatically evolves along a qubit-state-dependent loop in phase space. As a result, the two communication qubits acquire a nontrivial phase and become entangled.}
		\label{fig:RIPprinciple}
	\end{figure}

	The basic idea of the CIP gate is that by adiabatically turning on and off off-resonant drives, the state in the cavity traverses a qubit-state-dependent closed path in phase space~\cite{Dong2015_S,PhysRevLett.116.180501_S,PhysRevLett.117.250502_S}, as shown in Fig.~\ref{fig:RIPprinciple}. After this process, the cavity returns to the vacuum state, while different qubit states accumulate nontrivial phases.
	
	When we apply an off-resonant drive to cavity $S_2$, the Hamiltonian of the $Y_1-S_2-Y_2$ system in the rotating frame of the drive can be expressed as
	\begin{equation} 
		\begin{aligned}
			\hat H =&	(\delta+\frac{1}{2}\chi_{1}\hat\sigma_{Z_1}+\frac{1}{2}\chi_{2}\hat\sigma_{Z_2})\hat a^\dagger_{S_2}	\hat a_{S_2}
			+ \epsilon(t)(\hat a^\dagger_{S_2}	+\hat a_{S_2}),	
		\end{aligned}
		\label{eq:RIP Hamiltonian}
	\end{equation}
	where $\delta=\omega_{S_2}-\omega_{d}$, $\epsilon(t)$ is the time-dependent drive amplitude, $\chi_{1(2)}$ is the dispersive coupling strength between the communication qubit $Y_{1(2)}$ and cavity $S_2$, $\hat\sigma_{Z_1}$ ($\hat\sigma_{Z_2}$) is the Pauli-Z operator for the communication qubit, and $\hat a_{S_2}$ ($\hat a_{S_2}^\dagger$) is the annihilation (creation) operator for cavity $S_2$. 
	
	To get the effective two-qubit interaction Hamiltonian, we follow the derivation in Reference~\cite{PhysRevLett.116.180501_S}. We apply a time-dependent displacement to transform $\hat H_\mathrm{eff}=D(\hat\alpha')^\dagger \hat H D(\hat\alpha')-iD(\hat\alpha')^\dagger \dot D(\hat\alpha') $ to Hamiltonian ~\ref{eq:RIP Hamiltonian}. Here, $D(\hat\alpha')=\text{exp}(\hat\alpha'\hat a^\dagger-\hat\alpha'^*\hat a)$ and $\hat\alpha'=\alpha(1-\frac{\chi_{1}\hat\sigma_{Z_1}+\chi_{2}\hat\sigma_{Z_2}}{2\delta})$, where $\dot{\alpha}=-i\delta-i\epsilon(t)$. We can get the effective Hamiltonian:
	\begin{equation} 
		\begin{aligned}
			\hat H_\mathrm{eff} =&	\delta\hat a^\dagger_{S_2}		\hat a_{S_2}+\frac{1}{2}(\chi_{1}\hat\sigma_{Z_1}+\chi_{2}\hat\sigma_{Z_2})(\hat a^\dagger_{S_2}	\hat a_{S_2}+|\alpha|^2)\\
			&-\frac{\chi_{1}\chi_{2}|\alpha|^2}{2\delta}\hat\sigma_{Z_1}\hat\sigma_{Z_2}.	
		\end{aligned}
		\label{eq:Effective Hamiltonian}
	\end{equation}
	The nontrivial two-qubit phase originates from the last term of Hamiltonian~\ref{eq:Effective Hamiltonian}, and the rate of phase accumulation is proportional to ${|\alpha|^2}/{\delta}$. To ensure the adiabaticity of the process, we set a Gaussian profile for the CIP gate used in the experiment, defined as
	\begin{equation} \epsilon(t)=A\text{exp}(-\frac{8(t-\tau/2)^2}{\tau^2}).
		\label{eq:Gaussian}
	\end{equation}
	Here, $A$ is the maximum pump amplitude and $\tau$ is the total gate time. From Eq.~\ref{eq:Effective Hamiltonian} and Eq.~\ref{eq:Gaussian}, we can get $|\bar\alpha|\propto{A}/{\delta} $, and the rate of phase accumulation is then proportional to ${A^2}/{\delta^3} $. As a result, the total gate time $\tau\propto{\delta^3}/{A^2}$. Therefore, to shorten the gate time, we can either increase the pump amplitude $A$ or reduce $\delta$. However, it is important to note that an excessively large pump amplitude or a small detuning may destroy the adiabatic process, preventing the cavity from returning to the vacuum state.
	\subsection{Calibration of echoed CIP gate}
	\begin{figure}
		\includegraphics{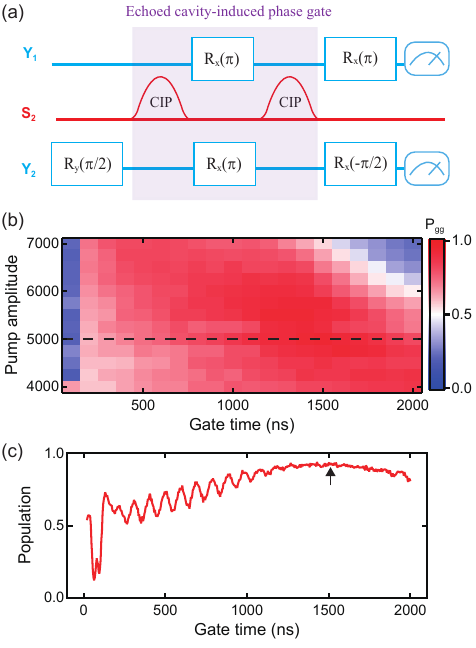}
		\caption{\textbf{Experimental sequence and results for the calibration of the echoed CIP gate.} (a) Experimental sequence for calibrating the echoed CIP gate. (b) Population of $\left|gg\right\rangle$ varies with the gate time (single CIP gate) and pump amplitude, with $\delta/2\pi=10~\text{MHz}$. (c) Linecut of (b) at the position indicated by the black dashed line. The gate time used in the experiment is indicated by the black array.}
		\label{fig:Echo}
	\end{figure}
	The superposition state of the two communication qubits $Y_1$ and $Y_2$, $\left|\psi_i\right\rangle=(\left|gg\right\rangle+\left|ge\right\rangle+\left|eg\right\rangle+\left|ee\right\rangle)/2$ will evolve into $\left|\psi_{f2}\right\rangle=(\left|gg\right\rangle+e^{i(\Phi_1+\Phi_2-\Phi_3)}\left|eg\right\rangle+e^{i(\Phi_1+\Phi_2-\Phi_3)}\left|ge\right\rangle+\left|ee\right\rangle)/2$ after the echoed CIP gate. To prepare the maximally entangled state, we need to satisfy the condition $\Phi_1+\Phi_2-\Phi_3=\pi/2$. 
	
	To determine the gate time required to meet such a condition, we design the experiment shown in Fig.~\ref{fig:Echo}(a). First, we prepare a superposition state on qubit $Y_2$ using an $R_y(\frac{\pi}{2})$ rotation pulse, and the joint state on $Y_1$ and $Y_2$ can be expressed as $\ket{\psi_0}=\frac{1}{\sqrt2}(\ket{gg}+\ket{ge})$. Next, we implement the echoed CIP gate, and $\ket{\psi_0}$ will undergo the following evolution: \begin{equation} 
		\begin{aligned}
			&\ket{\psi_0}=\frac{1}{\sqrt2}(\ket{gg}+\ket{ge})\\
			\xrightarrow {\mathrm{CIP}}&\ket{\psi_1}=\frac{1}{\sqrt2}(\ket{gg}+e^{i\Phi_1}\ket{ge})\\
			\xrightarrow {\mathrm{Echo}}&\ket{\psi_2}=\frac{1}{\sqrt2}(\ket{ee}+e^{i\Phi_1}\ket{eg})\\
			\xrightarrow {\mathrm{CIP}}&\ket{\psi_3}=\frac{1}{\sqrt2}(e^{i\Phi_3}\ket{ee}+e^{i(\Phi_1+\Phi_2)}\ket{eg}).\\
		\end{aligned}
		\label{eq:evolution}
	\end{equation}
	Finally, we apply an $R_x({\pi})$ rotation to qubit $Y_1$ and an $R_x(-\frac{\pi}{2})$ rotation to qubit $Y_2$. When $\Phi_1+\Phi_2-\Phi_3=\pi/2$, the state will come back to the ground state $\ket{gg}$. Conversely, if $\phi_1 + \phi_2 - \phi_3 \neq \pi/2$, the qubit state will not fully return to the ground state. In the experiment, we can scan the gate time, and the position that maximizes the population of $\ket{gg}$ corresponds to the required time.
	
	Figure~\ref{fig:Echo}(b) shows the measured population of $\ket{gg}$ ($P_{gg}$) as a function of pump amplitude and gate time with ${\delta}/{2\pi} = 10~\text{MHz}$. Figure~\ref{fig:Echo}(c) shows the linecut of Fig.~\ref{fig:Echo}(b) at the pump amplitude of $A=5000$. Here, the gate time represents the time of a single CIP gate. The data in Fig.~\ref{fig:Echo}(c) indicate a gate time of $\tau=1500$~ns, where $P_{gg}$ reaches its maximum value. It is noted that when the gate time is too short, $P_{gg}$ exhibits significant oscillations due to the rapid change rate of $\epsilon$, which breaks the adiabatic approximation. 
	
	To ensure that the parameters $({\delta}/{2\pi} = 10~\text{MHz},A=5000,\tau=1500~\text{ns})$ do not	result in excessive residual photons in $S_2$, we perform a numerical simulation following the sequence shown in Fig.~3(b) of the main text, using the decoherence parameters in Table~\ref{tab:SM_coherence}. The simulation results show that after the entire echoed CIP gate, the photon number in $S_2$ does not exceed 1\%. This confirms that these parameters satisfy the adiabatic condition well, and we finally adopt them for the experiment.
	\begin{table}
		\centering
		\caption{{Fidelities of the entangled communication-qubit states under different conditions in the storage cavities $S_1$ and $S_3$.}}
		\label{tab:Freentangle}
		\begin{tabular}{c|@{\hspace{10pt}} c} 
			\hline
			& $F_i$    \\ 
			\hline
			Both $S_1$ and $S_3$ are in vacuum (GRAPE)        & 0.963     \\
			Both $S_1$ and $S_3$ are in vacuum (Echoed CIP) & 0.931      	\\
			$S_1$ and $S_3$ are in \ket{01}+\ket{10} state (GRAPE)      & 0.923        \\
			$S_1$ and $S_3$ are in \ket{0_L1_L}+\ket{1_L0_L} state (Echoed CIP)          & 0.873      \\
			\hline
		\end{tabular}
	\end{table}
	{In Table~\ref{tab:Freentangle}, we summarize the fidelities of the entangled and re-entangled communication-qubit states implemented with GRAPE or echoed-CIP gates when storage cavities $S_1$ and $S_3$ are in different conditions.}

	\begin{figure}[b]
		\includegraphics{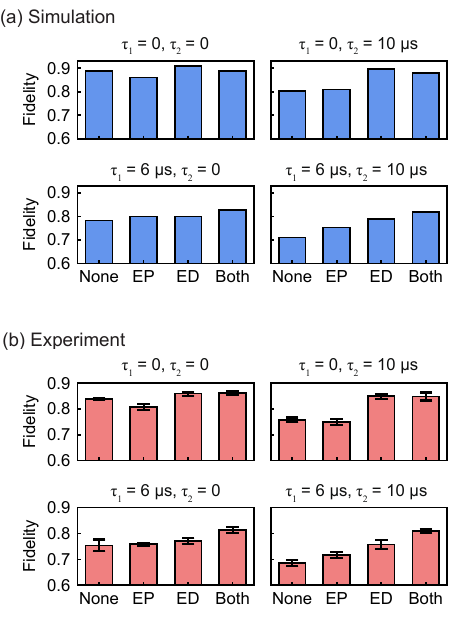}
		\caption{Comparison of simulated (a) and experimental (b) results for four purification schemes (no purification, EP only, ED only, and combination of both ED and EP) using the experimentally employed wait times ($\tau_1 = 0, 6~\mu$s and $\tau_2 = 0, 10~\mu$s).}
		\label{fig:S6}
	\end{figure}

	\section{Numerical simulations of EP$\&$ED for binomial codes}
\begin{figure}[t]
		\includegraphics{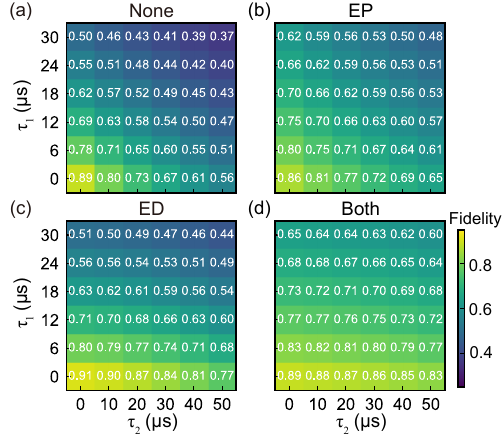}
		\caption{Simulation comparing the performance of four purification schemes under varying wait times $\tau_1$ and $\tau_2$: (a) no purification, (b) EP only, (c) ED only, and (d) combination of both ED and EP.}
		\label{fig:S7}
	\end{figure}

\begin{figure}[t]
		\includegraphics{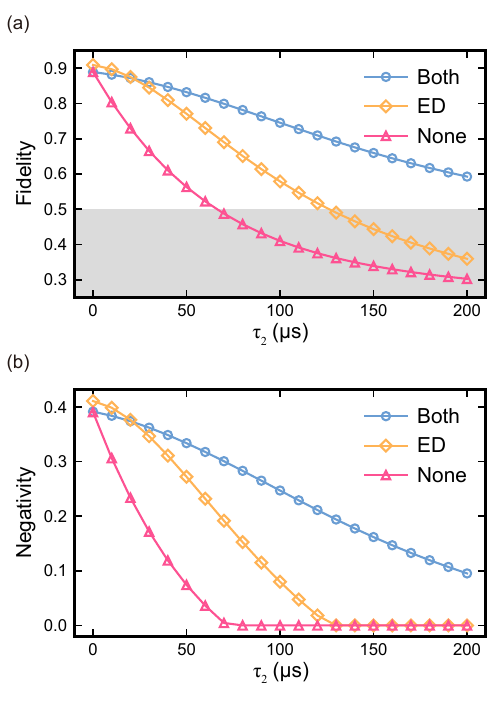}
		\caption{Simulated influence of different purification schemes on the existence time of entanglement in the logical qubits. The fidelity (a) and entanglement negativity (b) vary with $\tau_2$ for $\tau_1=0$.}
		\label{fig:S8}
	\end{figure}

	In the simulation, we use the coherence times and thermal populations listed in Table~\ref{tab:SM_coherence}, together with the experimentally measured fidelity of the re-entangled state [Fig.~3(c), right], to perform a full-process simulation of EP and ED. The simulation results, presented in Fig.~\ref{fig:S6} and Fig.~\ref{fig:S8}, show consistent behavior with the experimental data in Fig.~4(b) and Fig.~4(c). The offset between simulation and experiment arises from the measurement errors when photons are present in the cavities, as well as the fact that the GRAPE-optimized pulses typically achieve higher fidelity in simulation than in experiment because the GRAPE method uses tens of energy levels of the cavities in the optimization. Furthermore, to more clearly illustrate the effects of EP and ED beyond the experimentally accessible parameters and relaxation times, we perform additional simulation across the two relaxation-time dimensions ($\tau_1$ and $\tau_2$) for different purification schemes as shown in Fig.~\ref{fig:S7}.
	
	Based on the theoretical analysis and simulation results, we now analyze the data presented in Fig.~4(b). In the cases of $\tau_2=0,10\,\mathrm{\mu s}$, introducing $\tau_2$ primarily leads to unilateral single-photon-loss errors. In the cases of $\tau_1=0,6\,\mathrm{\mu s}$, increasing the relaxation time $\tau_1$ of the qubits enhances the population of $\ket{g}\ket{g}$ before encoding and the proportion of $\ket{0_L}\ket{0_L}$ after encoding, thereby introducing erasure errors.
	
	First, ED mitigates single-photon-loss errors arising from relaxation $\tau_2$, state preparation, and the $5.9\,\mathrm{\mu s}$ EP process. Consequently, applying ED generally improves the fidelity compared with the same conditions without ED. This effect is evident both when comparing ED alone with no operations and when comparing the combination of ED and EP with EP alone, as shown in Fig.~\ref{fig:S6} for both experimental and simulation results.

	Second, we compare the cases with no operations applied and with only EP applied. The theoretical analysis indicates that EP mitigates all two-photon-loss errors and $50\%$ of unilateral single-photon-loss errors. However, the $5.9\,\mathrm{\mu s}$ EP process itself can introduce additional single-photon loss and the measurement process introduces additional errors. Therefore, when no relaxation time is induced ($\tau_1=0,\tau_2=0$), the EP process can actually reduce the fidelity. When only $\tau_2$ is introduced, EP provides partial mitigation of single-photon-loss errors, leading to a smaller reduction in fidelity. When only $\tau_1$ is introduced, EP mitigates all erasure errors, resulting in a slight but not significant improvement in fidelity. When both $\tau_1$ and $\tau_2$ are present, EP mitigates both types of errors, producing the most significant increase in fidelity.
	
	Third, we consider the case when both EP and ED are applied simultaneously. EP mitigates all two-photon-loss errors and part of the single-photon-loss errors, but it also introduces some single-photon-loss errors. Subsequently applying ED suppresses all single-photon-loss errors. Therefore, as long as both two-photon-loss and single-photon-loss errors are significant in the system, the combined application of EP and ED outperforms all other scenarios.
	
	The error mitigation effects of different purification schemes, as illustrated in the two-dimensional simulation results of varying relaxation times $\tau_1$ and $\tau_2$ in Fig.~\ref{fig:S7}, more clearly confirm the conclusions discussed above. When $\tau_1$ is fixed, increasing $\tau_2$  gradually introduces more single-photon-loss errors. Both applying EP alone and ED alone significantly mitigate these errors, with the mitigation effect becoming stronger as $\tau_2$  increases. For single-photon-loss errors, EP can suppress all such errors, while ED has a theoretical limit of $50\%$, consistent with the observation in Fig.~\ref{fig:S7} that ED performs better than EP. When $\tau_2$ is fixed, increasing $\tau_1$ introduces more erasure errors, for which EP's mitigation effect improves, whereas ED's effectiveness decreases. Figure~\ref{fig:S7} also clearly shows that, except for a few cases near the origin, simultaneously applying EP and ED consistently yields the optimal results across other relaxation times.
	
	A more realistic scenario is illustrated in Fig.~\ref{fig:S8}, which shows the existence time of the entangled logical states stored in the two cavities. The coherence times of the two storage cavities are similar, $256\,\mathrm{\mu s}$ and $261\,\mathrm{\mu s}$, respectively. As shown in Fig.~\ref{fig:S7}, with increasing $\tau_2$, single-photon-loss errors become the dominant factor. Therefore, when no operations are applied, the entangled logical states disappear when a single-photon-loss error occurs. The ED process can mitigate single-photon loss on both sides, thereby suppressing errors to the occurrence of two-photon loss on a single side and extending the existence time. However, as $\tau_2$ increases, the proportion of two-photon-loss errors gradually becomes more significant. When EP and ED are applied simultaneously, unless both cavities experience two-photon loss, this strategy mitigates all other errors in the state. In theory, it can suppress errors down to the case where two-photon loss occurs in both cavities, thereby further increasing the existence time. Both simulation (Fig.~\ref{fig:S8}) and experimental results [Fig.~4(c)] for the entangled state existence time are consistent with this analysis, although practical factors such as measurement errors and imperfect gate operations result in shorter existence times than the theoretical expectation.
	
	Overall, based on our experimental results, theoretical analysis, and simulations, it is clear that ED effectively mitigates single-photon-loss errors, while EP mitigates two-photon-loss errors as well as part of the single-photon-loss errors. The combination of ED and EP can simultaneously and efficiently suppress both types of errors, and in theory, the ultimate limit is that the logical states stored in the two cavities can be preserved up to the point where only bilateral two-photon-loss errors remain.

\end{document}